\documentclass[aps,prb,reprint,showpacs,unsortedaddress]{revtex4-1}
\usepackage{amsmath}
\usepackage{amssymb}
\usepackage{graphicx}
\usepackage[normalem]{ulem}

\begin{document}

\title{Charge transfer and interfacial magnetism in (LaNiO$_3$)$_n$/(LaMnO$_3$)$_2$ superlattices}

\author{J. Hoffman}
\email{jhoffman@anl.gov}
\affiliation{Materials Science Division, Argonne National Laboratory, Argonne, Illinois 60439}

\author{I. C. Tung}
\affiliation{Department of Materials Science and Engineering, Northwestern University, Evanston, Illinois 60208}
\affiliation{Advanced Photon Source, Argonne National Laboratory, Argonne, Illinois 60439}

\author{B. B. Nelson-Cheeseman}
\affiliation{Materials Science Division, Argonne National Laboratory, Argonne, Illinois 60439}

\author{M. Liu}
\affiliation{Nanoscience and Technology Division, Argonne National Laboratory, Argonne, Illinois 60439}

\author{J. W. Freeland}
\affiliation{Advanced Photon Source, Argonne National Laboratory, Argonne, Illinois 60439}

\author{A. Bhattacharya}
\email{anand@anl.gov}
\affiliation{Materials Science Division, Argonne National Laboratory, Argonne, Illinois 60439}
\affiliation{Nanoscience and Technology Division, Argonne National Laboratory, Argonne, Illinois 60439}

\date{\today}

\begin{abstract}
   (LaNiO$_3$)$_n$/(LaMnO$_3$)$_2$ superlattices were grown using ozone-assisted molecular beam epitaxy, where LaNiO$_3$ is a paramagnetic metal and LaMnO$_3$ is an antiferromagnetic insulator.  The superlattices exhibit excellent crystallinity and interfacial roughness of less than 1 unit cell.  X-ray spectroscopy and dichroism measurements indicate that electrons are transferred from the LaMnO$_3$ to the LaNiO$_3$, inducing magnetism in LaNiO$_3$.  Magnetotransport measurements reveal a transition from metallic to insulating behavior as the LaNiO$_3$ layer thickness is reduced from 5 unit cells to 2 unit cells and suggest an inhomogeneous magnetic structure within LaNiO$_3$. 
\end{abstract}



\maketitle

\section{Introduction}
In recent years, there has been a great amount of interest in the novel electronic and magnetic states that emerge at interfaces between dissimilar complex oxide materials.\cite{HIK12} In the most commonly studied systems, these new behaviors arise as a result of interfacial charge redistribution and a resulting reconstruction of the orbital and spin degrees of freedom.  The interfacial charge redistribution can arise in several ways.  For example, at the widely studied LaAlO$_3$ (polar)/SrTiO$_3$ (non-polar) interface, this may result from a polar discontinuity.\cite{OH04, THS09, RTC07} Charge redistribution may also result from differences in chemical potential across an interface, as in LaTiO$_3$/SrTiO$_3$.\cite{OMG02a}  In magnetic systems, such as short-period LaMnO$_3$/SrMnO$_3$ superlattices, charge redistribution between neighboring Mn$^{3+}$/Mn$^{4+}$ sites can give rise to intermediate valency and a ferromagnetic, metallic ground state at the interface.\cite{KLF02, YKL06, BMV08}  Interfacial magnetism is linked to the charge state and bonding between the nearest neighbor $B$-site transition metal cations, and may therefore be very sensitive to the details of the structure.\cite{FCB10, MSK11}  Thus, creating tailored exchange interactions in oxide heterostructures requires precise control and understanding of the interface.

In this work, we examine the role of charge transfer in a series of digital superlattices that combine metallic LaNiO$_3$ (Ni$^{3+}$) with insulating antiferromagnetic LaMnO$_3$ (Mn$^{3+}$).  In this model system, each $B$O$_2$ atomic plane ($B$ = Mn, Ni) is sandwiched between two \emph{identical} $A$O (LaO) layers, thus ruling out intermixing of the $A$-side cation, which has been shown to lead to inadvertent doping.\cite{WPH07}  Through x-ray spectroscopy measurements we show the Mn to be in a 4+ oxidation state, while that of Ni is intermediate between 2+ and 3+, showing conclusively the presence of a charge transfer at the $\left[ 001 \right]$ LaNiO$_3$/LaMnO$_3$ interface.  The superlattices are found to have a net magnetic moment, with the magnetism residing on both the Mn and Ni sites.  We also observe a transition from metallic transport to insulating behavior as the thickness of the LNO layer is reduced below 4 unit cells.  These results, combined with magnetotransport measurements, point to the existence of a metallic state with inhomogeneous magnetism in LaNiO$_3$ with a mixture of Ni valence states.

\section{Materials and Experimental}
LaMnO$_3$ (LMO) has been extensively investigated as the parent compound of the prototypical perovskite colossal magnetoresistive (CMR) oxides.\cite{R97, CVV99} In stoichiometric bulk LMO with $t_{2g}^3e_g^1$ occupancy, a cooperative Jahn-Teller distortion lifts the degeneracy of the half-filled $e_g$ band, producing an orbitally-ordered $A$-type antiferromagnetic insulating ground state.\cite{AAB06} In thin films, however, LMO often exhibits ferromagnetism with a Curie temperature of $\sim$150 K and a saturation moment close to 4 $\mu_B$/Mn due to cation deficiency and strain.\cite{CMJ09, MSE10, GZS12} The LMO investigated here shows insulating behavior with a gap of $\sim$ 185 meV, a low saturation magnetization (M$_s <$ 0.6 $\mu_B$/Mn), and high coercivity, consistent with a weakly canted antiferromagnetic spin arrangement.  LaNiO$_3$ (LNO), on the other hand, is a paramagnetic metal where the Ni$^{3+}$ ion adopts a low spin, orbitally degenerate $t_{2g}^6e_g^1$ electronic configuration.  Strong mixing between the $d^7$ and $d^8\underline{L}$ ($\underline{L}$ denotes a ligand hole on the oxygen) configurations is expected, as LNO is significantly more covalent than divalent nickel compounds.  The introduction of Ni$^{2+}$ through oxygen deficiencies\cite{SCC96} or Ce-doping\cite{LGS05} has previously been shown to increase the resistivity of LNO and induce a transition to an insulating state.  Under our growth conditions, the resistivity of an 80 unit cell thick LNO film showed metallic behavior, with $\rho$(T = 5 K) = 35 $\mu\Omega$-cm, comparable to the lowest value for stoichiometric bulk samples.\cite{XPL93,GRX98}  

Epitaxial [(LaNiO$_3$)$_n$/(LaMnO$_3$)$_2$]$_m$ superlattices with $2 \leq n \leq 5$ unit cells were grown on (001) TiO$_2$-terminated SrTiO$_3$ single-crystal substrates using ozone assisted molecular beam epitaxy (MBE).  The total heterostructure thickness was kept constant at $\sim$80 unit cells ($\sim$30 nm) by adjusting the stacking periodicity, $m$.  Elemental materials were evaporated sequentially from effusion cells using a block-by-block technique described previously\cite{MSB09, SMR09} under an O$_3$ partial pressure of 2 $\times$ 10$^{-6}$ Torr, with the substrate maintained at 690$^{\circ}$C.  After growth, x-ray reflectivity (XRR) was used to confirm the thickness and determine the interfacial and surface roughnesses; the crystal structure and epitaxy were investigated with x-ray diffraction (XRD).  The in-plane resistivity was measured as a function of temperature in a four-point geometry for 2 K $\leq T \leq$ 300 K, while the Hall coefficient was measured using the Van der Pauw method in fields of up to 9 T.  Measurements of the net sample magnetization were carried out as a function of temperature and magnetic field using a superconducting quantum interference device (SQUID) magnetometer, while cation specific magnetic properties were investigated using x-ray magnetic circular dichroism (XMCD).  To determine the electronic properties, the superlattices were examined with x-ray absorption spectroscopy (XAS) in the soft x-ray regime at beamline 4-ID-C of the Advanced Photon Source (Argonne National Laboratory).  The spectra were measured in both the surface-sensitive total electron yield and bulk-sensitive fluorescence yield modes, and were aligned to a NiO (Ni$^{2+}$) standard measured simultaneously with the superlattice samples. 

\section{Results}
\begin{figure}[t]
  \includegraphics[width=\columnwidth]{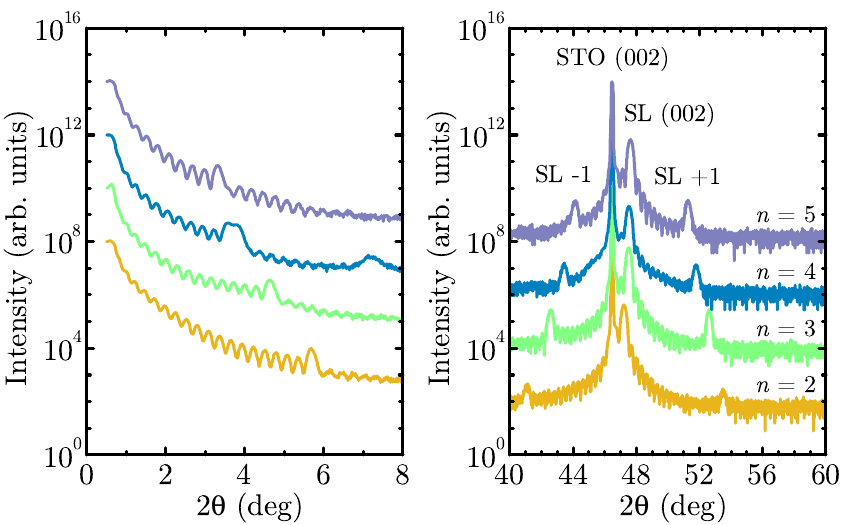}
	\caption{(Color online) X-ray reflectivity (a) and high-resolution x-ray diffraction (b) scans for a series of [(LaNiO$_3$)$_n$/(LaMnO$_3$)$_2$]$_m$ superlattices on SrTiO$_3$. The $c$-axis lattice parameters are found to be 0.3847 nm, 0.3824 nm, 0.3823 nm, and 0.3816 nm for $n=2,3,4,5$, respectively.}
	\label{fig:xrr_xrd}
\end{figure}
In order to obtain the intrinsic properties of the LMO/LNO interfaces, we require these to be atomically sharp with precise control of the stacking periodicity.  The thickness and lattice parameters of our samples were determined by x-ray reflectivity and high-resolution x-ray diffraction, as shown in Figs. \ref{fig:xrr_xrd}(a) and \ref{fig:xrr_xrd}(b), respectively.  Pronounced Bragg reflections indicate the interfaces are abrupt, as confirmed by fitting to the reflectivity curves using the Parratt formalism, which also showed the superlattice periods to typically be within 1\% of the nominal thicknesses.  We observe a ``double-peak'' structure in the XRR scan for the $n = 4$ sample, where the superlattice density modulation is not an exact integer number of unit cells, and for this sample estimate the error in stacking periodicity to be around 3\%.  The average $c$-axis lattice constant is obtained from $\theta-2\theta$ scans performed around the $(002)$ reflection as shown in Fig. \ref{fig:xrr_xrd}(b) and is found to decrease monotonically with increasing LNO thickness from 0.385 nm for $n$ = 2 to 0.382 nm when $n$ = 5,\footnote{The out-of-plane lattice constants for 80 unit cell thick films of LNO and LMO grown on (001) SrTiO$_3$ substrates were measured and found to be 0.394 nm and 0.381 nm, respectively.} in contrast with (LaNiO$_3$)$_n$/(SrMnO$_3$)$_2$ superlattices, where the largest out-of-plane lattice parameter was found for $n = 4$.\cite{MSB09}  The in-plane lattice parameters are determined by measurements of the $(103)$ and $(013)$ reflections and reveal that the superlattices are coherently clamped to the underlying SrTiO$_3$ substrate.

%
\begin{figure}[t]
  \includegraphics[width=\columnwidth]{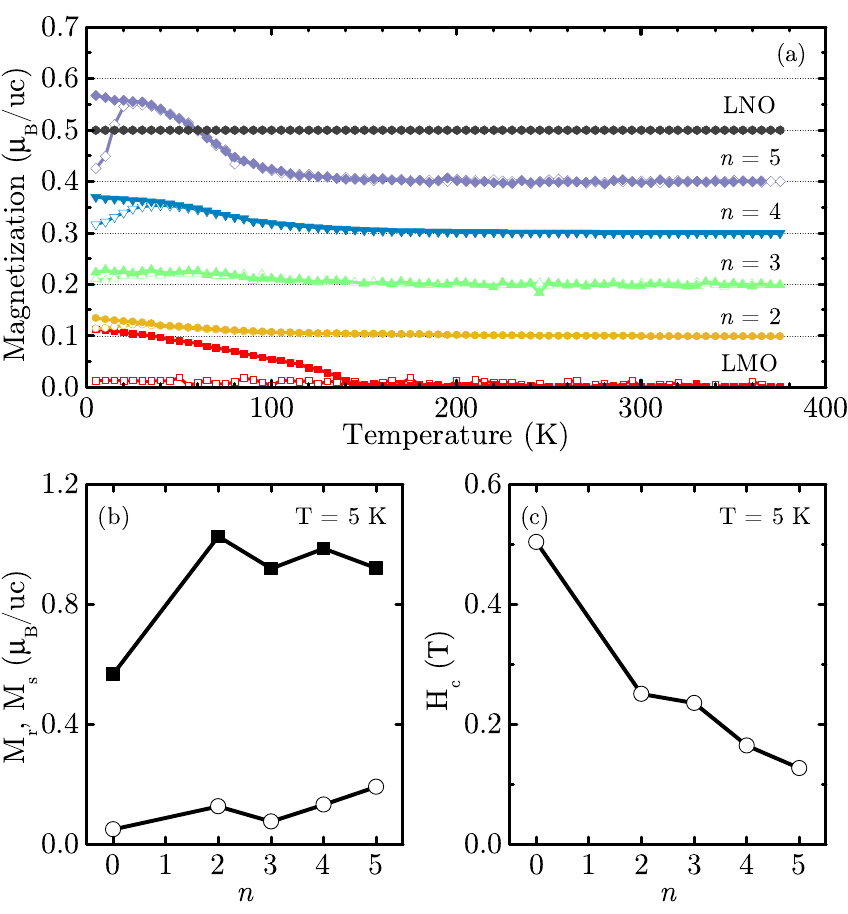}
	\caption{(Color online) (a) Temperature dependence of the magnetization for [(LaNiO$_3$)$_n$/(LaMnO$_3$)$_2$]$_m$ superlattices with $n$ = 2 -- 5 and 80 unit cell thick films of LMO and LNO.  Solid and open symbols show field-cooled and zero field-cooled measurements, respectively.  The field is applied in the plane of the film along the [100] direction.  (b) Variation with number of LNO layers of the remanent magnetization (open symbols) and saturation magnetization (filled symbols).  (c) Coercive field variation with number of LNO layers.}
	\label{fig:MvsT_MvsH}
\end{figure}
Figure \ref{fig:MvsT_MvsH}(a) shows the temperature dependence of the magnetization of the four superlattices in addition to LMO and LNO films.  The magnetization was measured while warming the samples in a field of 500 Oe applied in the plane of the film after field cooling in 500 Oe (solid symbols) and zero field cooling (open symbols).  On lowering the temperature, all of the samples except the LNO film, exhibit a rise in magnetization below a nominal transition temperature $T_C$ $\sim$150 K, corresponding to the onset of magnetic order.  From hysteresis loops measured along the [100] direction at 5 K (not shown) we determine the remanent and saturation magnetizations (Fig. \ref{fig:MvsT_MvsH}(b)), which show only weak variation for $2 \leq n \leq 5$.  The coercive field displayed a systematic reduction with increasing $n$, as shown in Fig. \ref{fig:MvsT_MvsH}(c).  The small saturation magnetization (M$_s <$ 0.6 $\mu_B$/Mn) and high coercivity ($\sim$5,000 Oe) of the pure LMO film are consistent with a weakly canted antiferromagnet ground state.

%
\begin{figure}[htbp]
 \includegraphics[width=\columnwidth]{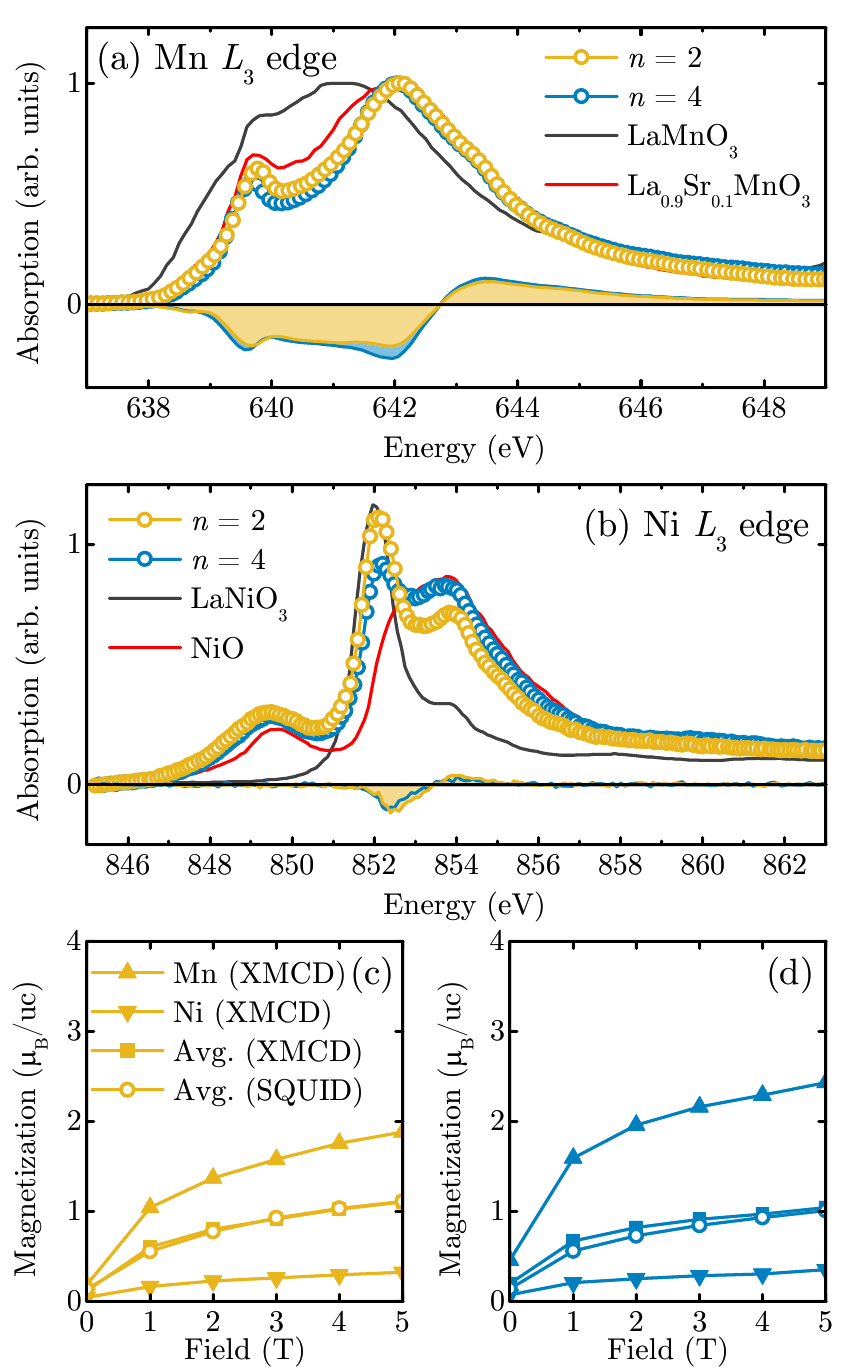}
	\caption{(Color online) (a) Mn L$_3$ and (b) Ni L$_3$ XAS and XMCD spectra for [(LaNiO$_3$)$_2$/(LaMnO$_3$)$_2$]$_{20}$ and [(LaNiO$_3$)$_4$/(LaMnO$_3$)$_2$]$_{13}$ superlattices. Mn$^{3+}$ and Mn$^{4+}$ references adapted from Ref. \onlinecite{ADF92}. (c) and (d) Cation-resolved magnetization as a function of field for the superlattices with $n = 2$ and $n= 4$, along with average magnetization (weighted by concentration of Mn and Ni) and magnetization obtained by SQUID magnetometry.}
	\label{fig:xas_xmcd}
\end{figure}
To probe the microscopic origins of the observed metal-insulator transition and magnetic ordering, we carried out both x-ray absorption spectroscopy (XAS) and x-ray magnetic circular dichroism (XMCD) measurements near the Mn and Ni $L_{2,3}$ absorption edges. Measurements were performed in fields of up to 5 T applied in the plane of the sample.  Both bulk-sensitive fluorescence yield and surface-sensitive total electron yield configurations were found to give similar results.  XAS spectra measured at T = 15 K are shown in Figs. \ref{fig:xas_xmcd}(a) and (b) for the superlattices with $n = 2$ and $n = 4$, along with spectra for Mn$^{3+}$ (LaMnO$_3$) and Mn$^{4+}$ (La$_{0.1}$Sr$_{0.9}$MnO$_3$) from Ref. \onlinecite{ADF92} and measured reference spectra for Ni$^{2+}$ (NiO) and Ni$^{3+}$ (LaNiO$_3$).  The peak at 849.5 eV in Fig. \ref{fig:xas_xmcd}(b) is due to the La $M_4$ transition.  We find that for both the $n = 2$ and $n = 4$ superlattices the valence of Mn is nearly Mn$^{4+}$, unlike in bulk LMO, where Mn is in the 3+ oxidation state.  The Ni valence is close to Ni$^{2+}$ in the $n = 2$ structure, while signatures of both Ni$^{2+}$ and Ni$^{3+}$ are found in the superlattice with 4 layers of LNO.  The Ni $L_3$-edge spectra of the $n = 4$ superlattice is similar to that reported for bulk $R$NiO$_3$ materials below the metal-insulator transition temperature\cite{M97} and for short-period LaNiO$_3$/LaAlO$_3$ superlattices,\cite{LOV11, FLK11} where the two-peak feature was attributed to the formation of charge-ordered states.  In our samples, we are unable to distinguish this scenario from a simple mixed valence picture without charge-ordering.  The results suggest that each Mn donates one electron to a nearby Ni cation at the interface, as in the double perovskite La$_2$NiMnO$_6$.\cite{GGV09}  The Ni$^{2+}$ is expected to have a doubly-degenerate, high-spin ($S$ = 1) $e_g$ manifold, with antiferromagnetic coupling to neighboring Ni$^{2+}$ cations, as in La$_2$NiO$_4$.  The competing exchange interactions at the interface may result in a frustrated magnetic state, as recently proposed to explain the exchange-bias effects observed in La$_{0.75}$Sr$_{0.25}$MnO$_3$/LaNiO$_3$ superlattices.\cite{SNG12}

In Figs. \ref{fig:xas_xmcd}(c) and (d) we show the net magnetization estimated for the Mn and Ni cations in the superlattices with $n = 2$ and $n = 4$, as the magnetic field is varied from 0.1 T to 5 T, with the temperature held constant at 15 K.\footnote{To quantify the magnetization of the individual cations, we compare the amplitude of the XMCD spectra with calibrated reference samples where the amplitude of the XMCD signal and magnetization are known.}  We find that the saturation Mn magnetization in both samples is \emph{enhanced} compared to the pure LMO film (Fig. \ref{fig:MvsT_MvsH}(b)), and the measured value of $\sim 2$ $\mu_B$/Mn is inconsistent with purely antiferromagnetic ordering of the Mn$^{4+}$ cations.  We note that Mn$^{4+}$ is expected to have an empty $e_g$ manifold (as in CaMnO$_3$), thus quenching the Jahn-Teller distortion and giving rise to $G$-type antiferromagnetic ordering.  The large magnetization that we observe may then occur by several mechanisms.  First, recent experimental work has shown that weak ferromagnetism can arise through phase separation in lightly electron-doped CaMnO$_3$.\cite{NC00}  Second, tensile strain, such as from coherent growth on SrTiO$_3$ substrates, has been shown theoretically to favor an $A$-type antiferromagnetic arrangement that may be susceptible to canting, leading to a net magnetization.\cite{TIT10}

We also find a net moment on the Ni cations, in agreement with Ref. \onlinecite{SNG12}, which we estimate to be 0.35 $\mu_B$/Ni at 5 T for both $n = 2$ and $n= 4$. From the shapes of the Mn and Ni $L_{2,3}$-edge spectra, we determine that the net moment of the LMO and LNO layers are coupled ferromagnetically.  To check the internal consistency of our XMCD magnetization estimates, we have computed the net magnetization of the superlattices by weighting the magnetization of the individual cations, as shown by the solid squares in Figs. \ref{fig:xas_xmcd}(c) and (d).  We find good agreement between the net magnetizations obtained from XMCD measurements and from SQUID magnetometry measurements (open circles in Figs. \ref{fig:xas_xmcd}(c) and (d)).\footnote{The diamagnetic contribution due to the SrTiO$_3$ substrate was measured and subtracted from the total magnetization measured by SQUID magnetometry.}

%
\begin{figure}[htbp]
  \includegraphics[width=\columnwidth]{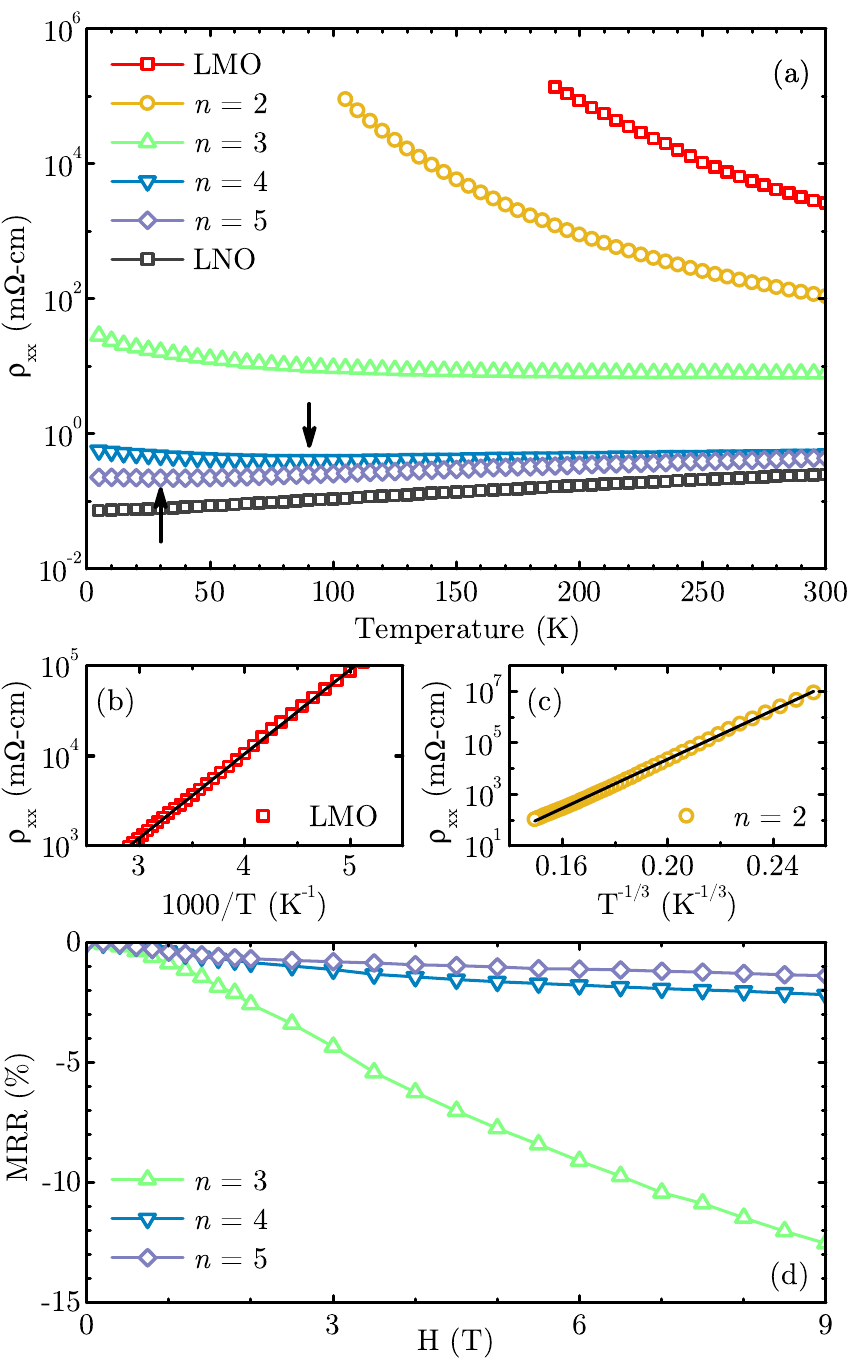}
	\caption{(Color online) Magnetotransport behavior of [(LaNiO$_3$)$_n$/(LaMnO$_3$)$_2$]$_m$ superlattices, $2 \leq n \leq 5$.  (a) Temperature dependence of longitudinal resistivity.  The arrows indicate positions of the resistivity minima at $T$ = 90 K $(n = 4)$ and $T$ = 30 K $(n = 5)$. (b) Resistivity of a pure LaMnO$_3$ film with a fit to thermally activated transport. (c) Fit to 2D variable-range hopping for $n = 2$ for 60 K $\leq T \leq $ 300 K. (d) Magnetoresistance ratio (MRR) of $n = 3$, $n = 4$, and $n = 5$ superlattices at T = 5 K with the field applied normal to the film.}
	\label{fig:transport_1}
\end{figure}
Figure \ref{fig:transport_1}(a) shows the variation of the resistivity of the LNO/LMO superlattices and LNO and LMO thin films as a function of temperature.  As shown in Fig. \ref{fig:transport_1}(b), the resistivity of an 80 unit cell LMO film shows thermally activated behavior, $\rho$ = $\rho_{\infty}\exp\left(E_A/k_BT\right)$ with an activation energy, $E_A = $185 meV.  The resistivity of the $n=2$ superlattice is well described by a two-dimensional variable-range hopping (VRH) model (Fig. \ref{fig:transport_1}(c)) given by,
\begin{equation}
\rho = \rho_\infty \exp\left[\left(T_0/T\right)^{1/3}\right],
\label{eq:VRH_2D}
\end{equation}
where $T_0$ is a characteristic temperature related to the density of states at the Fermi energy, $N(\varepsilon_F)$, and the localization length, $\xi$, by,\cite{SE84}
\begin{equation}
T_0 \approx \frac{13.8}{k_B N\left(\varepsilon_F\right)\xi^2}.
\label{eq:T0_2D}
\end{equation}
Two-dimensional VRH has previously been observed in ultrathin (5 unit cell) LNO films\cite{SGG11} and (LaNiO$_3$)$_2$/(SrMnO$_3$)$_2$ superlattices.\cite{MSB09} In the $n = 2$ sample, $T_0 \approx 1.31 \times 10^6$ K and the mean hopping energy $\Delta E \approx 0.28 k_B T_0^{1/3}T^{2/3}$, exceeds $k_BT$ over the entire temperature range studied ($60$ K $\leq T \leq 300$ K).  The density of states in bulk LNO is $\sim 1.1 \times 10^{29}$ eV$^{-1}$ m$^{-3}$,\cite{RSR91} from which we estimate $\xi \sim 0.07$ nm.  This localization length is incompatible with traditional theories of variable-range hopping where the localization length must be greater than the Ni-Ni distance and this may be due to the fact that the density of states we use is largely overestimated, consistent with our observation of Ni$^{2+}$ in this sample (Fig. \ref{fig:xas_xmcd}(b)).  This is similar to previous reports on hole-doped manganites,\cite{VRC97b} where a reduced $N\left(\varepsilon_F\right)$ was proposed to explain the insulating behavior.  This may occur due to the formation of a pseudogap.\cite{MYD99}

In previous studies of LaNiO$_3$/LaAlO$_3$,\cite{LOV11, BMB11} LaNiO$_3$/SrTiO$_3$,\cite{SLA10} and LaNiO$_3$/SrMnO$_3$\cite{MSB09} superlattices, a transition from metallic to insulating behavior was observed as the thickness of the LNO layer was reduced below 3 unit cells.  The fact that this transition occurs at the same thickness in the LNO/LMO superlattices studied here is surprising, as the average valence of the nickelate layer is lower due to charge transfer observed at the LMO/LNO interface.

To probe the origin of the thickness dependent metal to insulator transition, we performed magnetotransport measurements on the superlattices.  Figure \ref{fig:transport_1}(d) shows the magnetic field dependence of the magnetoresistance ratio (MRR) at $T = 5$ K for the superlattices with $n$ = 3 to 5.  (MRR is defined as $[R(H)-R(0)]/R(0)$.)  All of the superlattices show negative magnetoresistance, which is largest in the $n = 3$ sample, and decreases in the metallic samples.  The orientation dependence of the resistance (anisotropic magnetoresistance or AMR) was investigated in the $n = 4$ superlattice by rotating the sample about the [100]-axis in a fixed field of 9 T ($\mathbf{J}$ is along the [100] crystallographic direction and is maintained perpendicular to the magnetic field).  The resistance follows a $\cos^2 \theta$ dependence, where $\theta$ is the angle between the magnetic field and the normal to the film plane.  This is consistent with the AMR of ferromagnetic conductors, and further evidence that the carriers in LNO are scattering off magnetic moments in the superlattice.  

\begin{figure}[htbp]
  \includegraphics[width=\columnwidth]{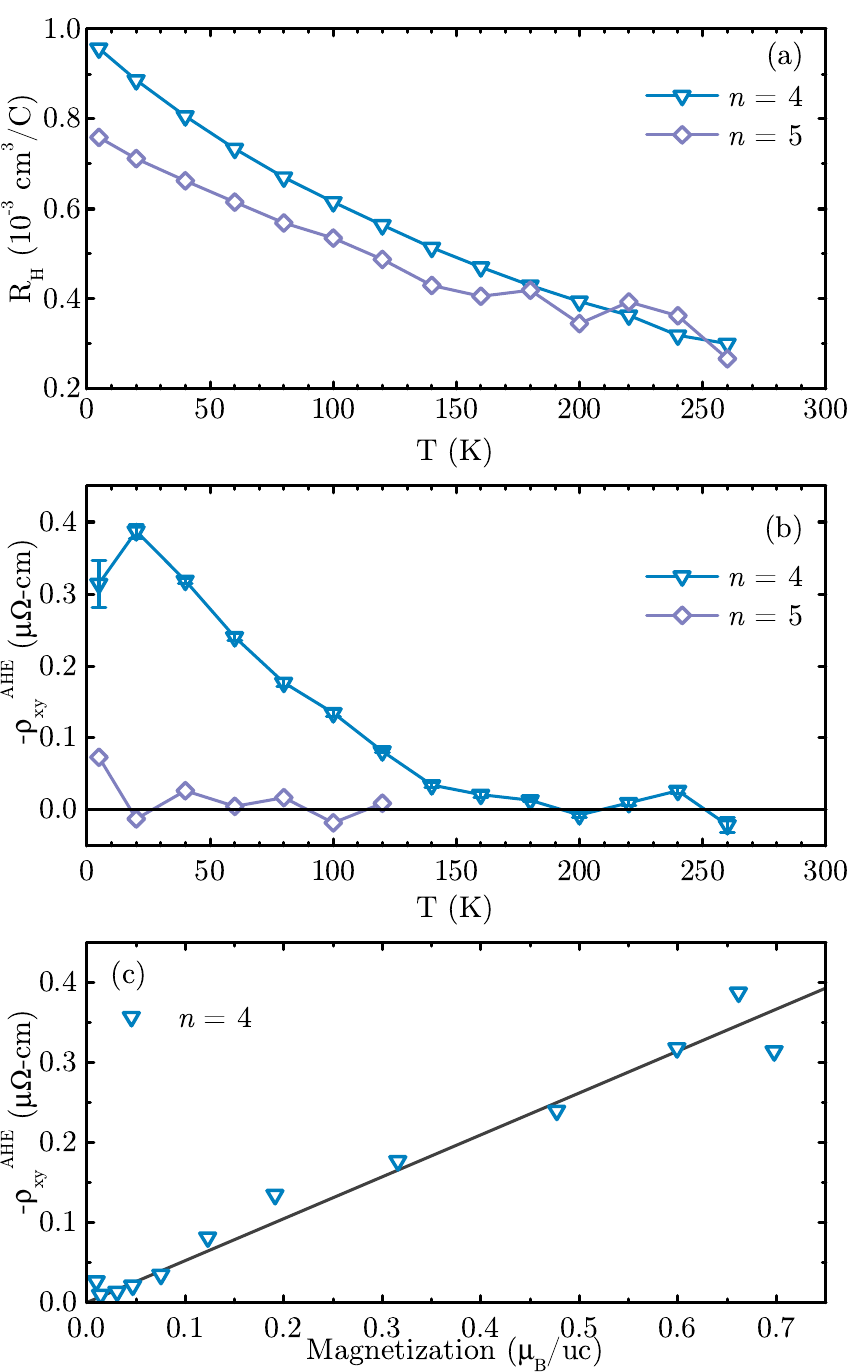}
	\caption{(Color online) (a) Ordinary Hall coefficient as a function of temperature for $n = 4$ and $n = 5$ superlattices. (b) Temperature dependence of the anomalous Hall resistivity for the $n = 4$ and $n = 5$ superlattices.  (c) Linear variation of the anomalous Hall resistivity with in-plane magnetization.}
	\label{fig:transport_2}
\end{figure}

In order to elucidate the nature of transport in the superlattices, we have measured the transverse (Hall) resistivity.  Empirically, for a ferromagnetic conductor this may be written as,
\begin{equation}
\rho_{xy} = R_H B + R_S \mu_0 M,
\label{eq:rho_xy}
\end{equation}
where $R_H$ is the ordinary Hall coefficient and $R_S$ is the anomalous Hall coefficient.  We determine $R_H$ from the high-field ($H > 4$ T) slope of the anti-symmetric transverse resistivity [i.e., ($\rho_{xy}(+H) - \rho_{xy}(-H))/2$], while the anomalous Hall resistivity, $\rho_{xy}^{AHE}(T) = R_S(T) \mu_0 M(T)$, is found by extrapolating the high-field slope to $H = 0$ T.  Figure \ref{fig:transport_2}(a) shows $R_H$ as a function of temperature for the metallic superlattices, where conduction is assumed to take place through the LNO layers only.  All of the samples show a positive Hall coefficient, consistent with hole-type conduction and in quantitative agreement with ultrathin LNO films\cite{SML10} and mixed-valence La$_{2-x}$Sr$_x$NiO$_4$ samples.\cite{SOK02}.  $R_H$ is found to decrease non-linearly with increasing temperature, which may point to the presence of two carrier types with different mobilities, as suggested by recent band structure calculations of LNO \cite{H93}.  The strong temperature dependence of $R_H$ has also been attributed to antiferromagnetic correlations in the metallic phase of systems on the verge of an insulator to metal transition, such as V$_{2-x}$O$_3$,\cite{CRM93} NiS$_{2-x}$Se$_x$,\cite{MTS00} and cuprate superconductors.\cite{NTS93, IKU93}  Assuming the current is confined to the LNO layers, we estimate the carrier concentration at $T = 5$ K for the metallic superlattices within a single-band model and find $p \approx 6.5 \times 10^{21}$ cm$^{-3}$ (0.38 holes per unit cell) and $p \approx 8.2 \times 10^{21} $cm$^{-3}$ (0.48 holes per unit cell) for samples with $n = 4$ and $n = 5$, respectively.  These values agree well with the expected number of holes per LNO unit cell, assuming each Ni cation at an interface donates a single hole to the neighboring Mn atom.

As shown in Fig. \ref{fig:transport_2}(b), we observe a negative anomalous Hall resistivity in the $n = 4$ sample below the magnetic transition temperature $T_C \sim$150 K.  The magnitude of the measured anomalous Hall resistivity is comparable to that previously observed in other ferromagnetic perovskite oxide systems.\cite{GWD00, YWH01,BZM04,NHH12}  In the $n = 5$ sample, however, $\rho_{xy}^{AHE}$ vanishes, despite the similarity of the bulk magnetic properties of the two superlattices  (Figs. \ref{fig:MvsT_MvsH}(b) and \ref{fig:MvsT_MvsH}(c)).  This unexpected behavior suggests the possibility of a inhomogeneous magnetic structure within the LNO layer, as was recently predicted to exist in $\langle 111 \rangle$-oriented LMO/LNO superlattices.\cite{GZS12}  In the $n = 4$ sample, conduction occurs in sufficiently close proximity to a magnetic interface to enhance spin-dependent scattering effects, giving rise to the observed anomalous Hall resistivity, while in the $n = 5$ sample, conduction takes place away from the spin-polarized interface through the middle of the LNO layer.  Furthermore, the superlattices considered here, in particular, those with low $n$, consist of dimensionally confined nickelate layers with intermediate valency, as found in the two-dimensional layered perovskite La$_{2-x}$Sr$_x$NiO$_4$.  Here, variation in the Ni oxidation state from Ni$^{2+}$ ($x$ = 0) to Ni$^{3+}$ ($x$ = 1) induce a series of changes in the electronic and magnetic ground state, with both stripe and checkerboard-type charge ordering/correlations observed at intermediate doping levels.\cite{SOK02}

The anomalous Hall effect arises from the spin-orbit interaction in the presence of broken time-reversal symmetry.\cite{NSO10}  In homogeneous magnetic systems, $R_S$ is predicted to scale with the longitudinal resistivity: $R_S \propto \rho_{xx}^{\gamma}$, with $\gamma = 1$ for skew-scattering and $\gamma = 2$ for the quantum mechanical side-jump mechanism.  In magnetically inhomogeneous systems, such as Co/Pt superlattices\cite{CLX00} and granular Co-Ag,\cite{XXW92} values of $\gamma$ greater than 2 have been reported.  We analyze the anomalous Hall effect in the $n = 4$ sample qualitatively by plotting in Fig. \ref{fig:transport_2}(c) $\rho_{xy}^{AHE}(T)$ as function of $M(T)$, the in-plane magnetization of the superlattice measured in a field of 500 Oe after field cooling.  The linear dependence between $\rho_{xy}^{AHE}(T)$ and $M(T)$ suggests that $R_S$ is a constant, independent of temperature.  However, as shown in Fig. \ref{fig:transport_1}(a), $\rho_{xx}$ of the $n = 4$ superlattice has a minimum at $T = 90$ K, ruling out a simple power law scaling relation between $R_S$ and $\rho_{xx}$.  A lack of scaling between $R_S$ and $\rho_{xx}$ has also been observed in colossal magnetoresistive manganites, where $R_S$ is sharply peaked above the Curie temperature.  This behavior is attributed to Berry phase effects arising from fluctuating non-collinear spin textures.\cite{YKM99}  This is, however, in contrast to our samples, where $R_S$ does not depend on temperature, which may be due to a frozen interfacial spin texture, though a quantitative model is lacking.  Quantitative determination of $R_S$ is beyond the scope of this paper, and requires knowledge of the LNO magnetization as functions of both applied field and temperature.

\section{Conclusions}
In conclusion, detailed magnetotransport, magnetic, and spectroscopic measurements were carried out on atomically sharp [001] (LaNiO$_3$)$_n$/(LaMnO$_3$)$_2$ superlattices.  We have unambiguous evidence of the transfer of electrons from Mn to Ni, and ferromagnetic coupling of the net magnetization on the Mn and Ni sites.  As the LNO layer thickness increases, the average Ni valence changes from Ni$^{2+}$ ($n = 2$) to Ni$^{2.5+}$ ($n = 4$), and is accompanied by a transition from insulating behavior to metallic transport.  Detailed magnetotransport measurements suggest the carriers in the LaNiO$_3$ layer scatter off the magnetic Ni sites at the interface, and that the magnetization on the Ni is inhomogeneous.  This work opens new avenues for the creation of artificial oxide heterostructures with tailored electronic and magnetic properties.   

\section{Acknowledgments}
Work at Argonne National Laboratory, including the use of the Center for Nanoscale Materials, supported by the U.S. Department of Energy, Office of Basic Energy Sciences under contract no. DE-AC02-06CH11357.

\bibliography{../../library}

\begin{thebibliography}{56}%
\makeatletter
\providecommand \@ifxundefined [1]{%
 \@ifx{#1\undefined}
}%
\providecommand \@ifnum [1]{%
 \ifnum #1\expandafter \@firstoftwo
 \else \expandafter \@secondoftwo
 \fi
}%
\providecommand \@ifx [1]{%
 \ifx #1\expandafter \@firstoftwo
 \else \expandafter \@secondoftwo
 \fi
}%
\providecommand \natexlab [1]{#1}%
\providecommand \enquote  [1]{``#1''}%
\providecommand \bibnamefont  [1]{#1}%
\providecommand \bibfnamefont [1]{#1}%
\providecommand \citenamefont [1]{#1}%
\providecommand \href@noop [0]{\@secondoftwo}%
\providecommand \href [0]{\begingroup \@sanitize@url \@href}%
\providecommand \@href[1]{\@@startlink{#1}\@@href}%
\providecommand \@@href[1]{\endgroup#1\@@endlink}%
\providecommand \@sanitize@url [0]{\catcode `\\12\catcode `\$12\catcode
  `\&12\catcode `\#12\catcode `\^12\catcode `\_12\catcode `\%12\relax}%
\providecommand \@@startlink[1]{}%
\providecommand \@@endlink[0]{}%
\providecommand \url  [0]{\begingroup\@sanitize@url \@url }%
\providecommand \@url [1]{\endgroup\@href {#1}{\urlprefix }}%
\providecommand \urlprefix  [0]{URL }%
\providecommand \Eprint [0]{\href }%
\providecommand \doibase [0]{http://dx.doi.org/}%
\providecommand \selectlanguage [0]{\@gobble}%
\providecommand \bibinfo  [0]{\@secondoftwo}%
\providecommand \bibfield  [0]{\@secondoftwo}%
\providecommand \translation [1]{[#1]}%
\providecommand \BibitemOpen [0]{}%
\providecommand \bibitemStop [0]{}%
\providecommand \bibitemNoStop [0]{.\EOS\space}%
\providecommand \EOS [0]{\spacefactor3000\relax}%
\providecommand \BibitemShut  [1]{\csname bibitem#1\endcsname}%
\let\auto@bib@innerbib\@empty
\bibitem [{\citenamefont {Hwang}\ \emph {et~al.}(2012)\citenamefont {Hwang},
  \citenamefont {Iwasa}, \citenamefont {Kawasaki}, \citenamefont {Keimer},
  \citenamefont {Nagaosa},\ and\ \citenamefont {Tokura}}]{HIK12}%
  \BibitemOpen
  \bibfield  {author} {\bibinfo {author} {\bibfnamefont {H.~Y.}\ \bibnamefont
  {Hwang}}, \bibinfo {author} {\bibfnamefont {Y.}~\bibnamefont {Iwasa}},
  \bibinfo {author} {\bibfnamefont {M.}~\bibnamefont {Kawasaki}}, \bibinfo
  {author} {\bibfnamefont {B.}~\bibnamefont {Keimer}}, \bibinfo {author}
  {\bibfnamefont {N.}~\bibnamefont {Nagaosa}}, \ and\ \bibinfo {author}
  {\bibfnamefont {Y.}~\bibnamefont {Tokura}},\ }\href {\doibase
  10.1038/nmat3223} {\bibfield  {journal} {\bibinfo  {journal} {Nature
  Materials}\ }\textbf {\bibinfo {volume} {11}},\ \bibinfo {pages} {103}
  (\bibinfo {year} {2012})}\BibitemShut {NoStop}%
\bibitem [{\citenamefont {Ohtomo}\ and\ \citenamefont {Hwang}(2004)}]{OH04}%
  \BibitemOpen
  \bibfield  {author} {\bibinfo {author} {\bibfnamefont {A.}~\bibnamefont
  {Ohtomo}}\ and\ \bibinfo {author} {\bibfnamefont {H.~Y.}\ \bibnamefont
  {Hwang}},\ }\href {\doibase 10.1038/nature02308} {\bibfield  {journal}
  {\bibinfo  {journal} {Nature}\ }\textbf {\bibinfo {volume} {427}},\ \bibinfo
  {pages} {423} (\bibinfo {year} {2004})}\BibitemShut {NoStop}%
\bibitem [{\citenamefont {Takizawa}\ \emph {et~al.}(2009)\citenamefont
  {Takizawa}, \citenamefont {Hotta}, \citenamefont {Susaki}, \citenamefont
  {Ishida}, \citenamefont {Wadati}, \citenamefont {Takata}, \citenamefont
  {Horiba}, \citenamefont {Matsunami}, \citenamefont {Shin}, \citenamefont
  {Yabashi}, \citenamefont {Tamasaku}, \citenamefont {Nishino}, \citenamefont
  {Ishikawa}, \citenamefont {Fujimori},\ and\ \citenamefont {Hwang}}]{THS09}%
  \BibitemOpen
  \bibfield  {author} {\bibinfo {author} {\bibfnamefont {M.}~\bibnamefont
  {Takizawa}}, \bibinfo {author} {\bibfnamefont {Y.}~\bibnamefont {Hotta}},
  \bibinfo {author} {\bibfnamefont {T.}~\bibnamefont {Susaki}}, \bibinfo
  {author} {\bibfnamefont {Y.}~\bibnamefont {Ishida}}, \bibinfo {author}
  {\bibfnamefont {H.}~\bibnamefont {Wadati}}, \bibinfo {author} {\bibfnamefont
  {Y.}~\bibnamefont {Takata}}, \bibinfo {author} {\bibfnamefont
  {K.}~\bibnamefont {Horiba}}, \bibinfo {author} {\bibfnamefont
  {M.}~\bibnamefont {Matsunami}}, \bibinfo {author} {\bibfnamefont
  {S.}~\bibnamefont {Shin}}, \bibinfo {author} {\bibfnamefont {M.}~\bibnamefont
  {Yabashi}}, \bibinfo {author} {\bibfnamefont {K.}~\bibnamefont {Tamasaku}},
  \bibinfo {author} {\bibfnamefont {Y.}~\bibnamefont {Nishino}}, \bibinfo
  {author} {\bibfnamefont {T.}~\bibnamefont {Ishikawa}}, \bibinfo {author}
  {\bibfnamefont {A.}~\bibnamefont {Fujimori}}, \ and\ \bibinfo {author}
  {\bibfnamefont {H.}~\bibnamefont {Hwang}},\ }\href {\doibase
  10.1103/PhysRevLett.102.236401} {\bibfield  {journal} {\bibinfo  {journal}
  {Physical Review Letters}\ }\textbf {\bibinfo {volume} {102}},\ \bibinfo
  {pages} {236401} (\bibinfo {year} {2009})}\BibitemShut {NoStop}%
\bibitem [{\citenamefont {Reyren}\ \emph {et~al.}(2007)\citenamefont {Reyren},
  \citenamefont {Thiel}, \citenamefont {Caviglia}, \citenamefont {Kourkoutis},
  \citenamefont {Hammerl}, \citenamefont {Richter}, \citenamefont {Schneider},
  \citenamefont {Kopp}, \citenamefont {R\"{u}etschi}, \citenamefont {Jaccard},
  \citenamefont {Gabay}, \citenamefont {Muller}, \citenamefont {Triscone},\
  and\ \citenamefont {Mannhart}}]{RTC07}%
  \BibitemOpen
  \bibfield  {author} {\bibinfo {author} {\bibfnamefont {N.}~\bibnamefont
  {Reyren}}, \bibinfo {author} {\bibfnamefont {S.}~\bibnamefont {Thiel}},
  \bibinfo {author} {\bibfnamefont {A.~D.}\ \bibnamefont {Caviglia}}, \bibinfo
  {author} {\bibfnamefont {L.~F.}\ \bibnamefont {Kourkoutis}}, \bibinfo
  {author} {\bibfnamefont {G.}~\bibnamefont {Hammerl}}, \bibinfo {author}
  {\bibfnamefont {C.}~\bibnamefont {Richter}}, \bibinfo {author} {\bibfnamefont
  {C.~W.}\ \bibnamefont {Schneider}}, \bibinfo {author} {\bibfnamefont
  {T.}~\bibnamefont {Kopp}}, \bibinfo {author} {\bibfnamefont {A.-S.}\
  \bibnamefont {R\"{u}etschi}}, \bibinfo {author} {\bibfnamefont
  {D.}~\bibnamefont {Jaccard}}, \bibinfo {author} {\bibfnamefont
  {M.}~\bibnamefont {Gabay}}, \bibinfo {author} {\bibfnamefont {D.~A.}\
  \bibnamefont {Muller}}, \bibinfo {author} {\bibfnamefont {J.-M.}\
  \bibnamefont {Triscone}}, \ and\ \bibinfo {author} {\bibfnamefont
  {J.}~\bibnamefont {Mannhart}},\ }\href {\doibase 10.1126/science.1146006}
  {\bibfield  {journal} {\bibinfo  {journal} {Science}\ }\textbf {\bibinfo
  {volume} {317}},\ \bibinfo {pages} {1196} (\bibinfo {year}
  {2007})}\BibitemShut {NoStop}%
\bibitem [{\citenamefont {Ohtomo}\ \emph {et~al.}(2002)\citenamefont {Ohtomo},
  \citenamefont {Muller}, \citenamefont {Grazul},\ and\ \citenamefont
  {Hwang}}]{OMG02a}%
  \BibitemOpen
  \bibfield  {author} {\bibinfo {author} {\bibfnamefont {A.}~\bibnamefont
  {Ohtomo}}, \bibinfo {author} {\bibfnamefont {D.~A.}\ \bibnamefont {Muller}},
  \bibinfo {author} {\bibfnamefont {J.~L.}\ \bibnamefont {Grazul}}, \ and\
  \bibinfo {author} {\bibfnamefont {H.~Y.}\ \bibnamefont {Hwang}},\ }\href
  {\doibase 10.1038/nature00977} {\bibfield  {journal} {\bibinfo  {journal}
  {Nature}\ }\textbf {\bibinfo {volume} {419}},\ \bibinfo {pages} {378}
  (\bibinfo {year} {2002})}\BibitemShut {NoStop}%
\bibitem [{\citenamefont {Koida}\ \emph {et~al.}(2002)\citenamefont {Koida},
  \citenamefont {Lippmaa}, \citenamefont {Fukumura}, \citenamefont {Itaka},
  \citenamefont {Matsumoto}, \citenamefont {Kawasaki},\ and\ \citenamefont
  {Koinuma}}]{KLF02}%
  \BibitemOpen
  \bibfield  {author} {\bibinfo {author} {\bibfnamefont {T.}~\bibnamefont
  {Koida}}, \bibinfo {author} {\bibfnamefont {M.}~\bibnamefont {Lippmaa}},
  \bibinfo {author} {\bibfnamefont {T.}~\bibnamefont {Fukumura}}, \bibinfo
  {author} {\bibfnamefont {K.}~\bibnamefont {Itaka}}, \bibinfo {author}
  {\bibfnamefont {Y.}~\bibnamefont {Matsumoto}}, \bibinfo {author}
  {\bibfnamefont {M.}~\bibnamefont {Kawasaki}}, \ and\ \bibinfo {author}
  {\bibfnamefont {H.}~\bibnamefont {Koinuma}},\ }\href {\doibase
  10.1103/PhysRevB.66.144418} {\bibfield  {journal} {\bibinfo  {journal}
  {Physical Review B}\ }\textbf {\bibinfo {volume} {66}},\ \bibinfo {pages}
  {144418} (\bibinfo {year} {2002})}\BibitemShut {NoStop}%
\bibitem [{\citenamefont {Yamada}\ \emph {et~al.}(2006)\citenamefont {Yamada},
  \citenamefont {Kawasaki}, \citenamefont {Lottermoser}, \citenamefont
  {Arima},\ and\ \citenamefont {Tokura}}]{YKL06}%
  \BibitemOpen
  \bibfield  {author} {\bibinfo {author} {\bibfnamefont {H.}~\bibnamefont
  {Yamada}}, \bibinfo {author} {\bibfnamefont {M.}~\bibnamefont {Kawasaki}},
  \bibinfo {author} {\bibfnamefont {T.}~\bibnamefont {Lottermoser}}, \bibinfo
  {author} {\bibfnamefont {T.}~\bibnamefont {Arima}}, \ and\ \bibinfo {author}
  {\bibfnamefont {Y.}~\bibnamefont {Tokura}},\ }\href {\doibase
  10.1063/1.2266863} {\bibfield  {journal} {\bibinfo  {journal} {Applied
  Physics Letters}\ }\textbf {\bibinfo {volume} {89}},\ \bibinfo {pages}
  {52506} (\bibinfo {year} {2006})}\BibitemShut {NoStop}%
\bibitem [{\citenamefont {Bhattacharya}\ \emph {et~al.}(2008)\citenamefont
  {Bhattacharya}, \citenamefont {May}, \citenamefont {te~Velthuis},
  \citenamefont {Warusawithana}, \citenamefont {Zhai}, \citenamefont {Jiang},
  \citenamefont {Zuo}, \citenamefont {Fitzsimmons}, \citenamefont {Bader},\
  and\ \citenamefont {Eckstein}}]{BMV08}%
  \BibitemOpen
  \bibfield  {author} {\bibinfo {author} {\bibfnamefont {A.}~\bibnamefont
  {Bhattacharya}}, \bibinfo {author} {\bibfnamefont {S.~J.}\ \bibnamefont
  {May}}, \bibinfo {author} {\bibfnamefont {S.~G.~E.}\ \bibnamefont
  {te~Velthuis}}, \bibinfo {author} {\bibfnamefont {M.}~\bibnamefont
  {Warusawithana}}, \bibinfo {author} {\bibfnamefont {X.}~\bibnamefont {Zhai}},
  \bibinfo {author} {\bibfnamefont {B.}~\bibnamefont {Jiang}}, \bibinfo
  {author} {\bibfnamefont {J.-M.}\ \bibnamefont {Zuo}}, \bibinfo {author}
  {\bibfnamefont {M.~R.}\ \bibnamefont {Fitzsimmons}}, \bibinfo {author}
  {\bibfnamefont {S.~D.}\ \bibnamefont {Bader}}, \ and\ \bibinfo {author}
  {\bibfnamefont {J.~N.}\ \bibnamefont {Eckstein}},\ }\href {\doibase
  10.1103/PhysRevLett.100.257203} {\bibfield  {journal} {\bibinfo  {journal}
  {Physical Review Letters}\ }\textbf {\bibinfo {volume} {100}},\ \bibinfo
  {pages} {257203} (\bibinfo {year} {2008})}\BibitemShut {NoStop}%
\bibitem [{\citenamefont {Freeland}\ \emph {et~al.}(2010)\citenamefont
  {Freeland}, \citenamefont {Chakhalian}, \citenamefont {Boris}, \citenamefont
  {Tonnerre}, \citenamefont {Kavich}, \citenamefont {Yordanov}, \citenamefont
  {Grenier}, \citenamefont {Zschack}, \citenamefont {Karapetrova},
  \citenamefont {Popovich}, \citenamefont {Lee},\ and\ \citenamefont
  {Keimer}}]{FCB10}%
  \BibitemOpen
  \bibfield  {author} {\bibinfo {author} {\bibfnamefont {J.~W.}\ \bibnamefont
  {Freeland}}, \bibinfo {author} {\bibfnamefont {J.}~\bibnamefont
  {Chakhalian}}, \bibinfo {author} {\bibfnamefont {A.~V.}\ \bibnamefont
  {Boris}}, \bibinfo {author} {\bibfnamefont {J.~M.}\ \bibnamefont {Tonnerre}},
  \bibinfo {author} {\bibfnamefont {J.~J.}\ \bibnamefont {Kavich}}, \bibinfo
  {author} {\bibfnamefont {P.}~\bibnamefont {Yordanov}}, \bibinfo {author}
  {\bibfnamefont {S.}~\bibnamefont {Grenier}}, \bibinfo {author} {\bibfnamefont
  {P.}~\bibnamefont {Zschack}}, \bibinfo {author} {\bibfnamefont
  {E.}~\bibnamefont {Karapetrova}}, \bibinfo {author} {\bibfnamefont
  {P.}~\bibnamefont {Popovich}}, \bibinfo {author} {\bibfnamefont {H.~N.}\
  \bibnamefont {Lee}}, \ and\ \bibinfo {author} {\bibfnamefont
  {B.}~\bibnamefont {Keimer}},\ }\href {\doibase 10.1103/PhysRevB.81.094414}
  {\bibfield  {journal} {\bibinfo  {journal} {Physical Review B}\ }\textbf
  {\bibinfo {volume} {81}},\ \bibinfo {pages} {94414} (\bibinfo {year}
  {2010})}\BibitemShut {NoStop}%
\bibitem [{\citenamefont {May}\ \emph {et~al.}(2011)\citenamefont {May},
  \citenamefont {Smith}, \citenamefont {Kim}, \citenamefont {Karapetrova},
  \citenamefont {Bhattacharya},\ and\ \citenamefont {Ryan}}]{MSK11}%
  \BibitemOpen
  \bibfield  {author} {\bibinfo {author} {\bibfnamefont {S.~J.}\ \bibnamefont
  {May}}, \bibinfo {author} {\bibfnamefont {C.~R.}\ \bibnamefont {Smith}},
  \bibinfo {author} {\bibfnamefont {J.-W.}\ \bibnamefont {Kim}}, \bibinfo
  {author} {\bibfnamefont {E.}~\bibnamefont {Karapetrova}}, \bibinfo {author}
  {\bibfnamefont {A.}~\bibnamefont {Bhattacharya}}, \ and\ \bibinfo {author}
  {\bibfnamefont {P.~J.}\ \bibnamefont {Ryan}},\ }\href {\doibase
  10.1103/PhysRevB.83.153411} {\bibfield  {journal} {\bibinfo  {journal}
  {Physical Review B}\ }\textbf {\bibinfo {volume} {83}},\ \bibinfo {pages}
  {153411} (\bibinfo {year} {2011})}\BibitemShut {NoStop}%
\bibitem [{\citenamefont {Willmott}\ \emph {et~al.}(2007)\citenamefont
  {Willmott}, \citenamefont {Pauli}, \citenamefont {Herger}, \citenamefont
  {Schlep\"{u}tz}, \citenamefont {Martoccia}, \citenamefont {Patterson},
  \citenamefont {Delley}, \citenamefont {Clarke}, \citenamefont {Kumah},
  \citenamefont {Cionca},\ and\ \citenamefont {Yacoby}}]{WPH07}%
  \BibitemOpen
  \bibfield  {author} {\bibinfo {author} {\bibfnamefont {P.}~\bibnamefont
  {Willmott}}, \bibinfo {author} {\bibfnamefont {S.}~\bibnamefont {Pauli}},
  \bibinfo {author} {\bibfnamefont {R.}~\bibnamefont {Herger}}, \bibinfo
  {author} {\bibfnamefont {C.}~\bibnamefont {Schlep\"{u}tz}}, \bibinfo {author}
  {\bibfnamefont {D.}~\bibnamefont {Martoccia}}, \bibinfo {author}
  {\bibfnamefont {B.}~\bibnamefont {Patterson}}, \bibinfo {author}
  {\bibfnamefont {B.}~\bibnamefont {Delley}}, \bibinfo {author} {\bibfnamefont
  {R.}~\bibnamefont {Clarke}}, \bibinfo {author} {\bibfnamefont
  {D.}~\bibnamefont {Kumah}}, \bibinfo {author} {\bibfnamefont
  {C.}~\bibnamefont {Cionca}}, \ and\ \bibinfo {author} {\bibfnamefont
  {Y.}~\bibnamefont {Yacoby}},\ }\href {\doibase 10.1103/PhysRevLett.99.155502}
  {\bibfield  {journal} {\bibinfo  {journal} {Physical Review Letters}\
  }\textbf {\bibinfo {volume} {99}},\ \bibinfo {pages} {155502} (\bibinfo
  {year} {2007})}\BibitemShut {NoStop}%
\bibitem [{\citenamefont {Ramirez}(1997)}]{R97}%
  \BibitemOpen
  \bibfield  {author} {\bibinfo {author} {\bibfnamefont {A.~P.}\ \bibnamefont
  {Ramirez}},\ }\href {\doibase 10.1088/0953-8984/9/39/005} {\bibfield
  {journal} {\bibinfo  {journal} {Journal of Physics: Condensed Matter}\
  }\textbf {\bibinfo {volume} {9}},\ \bibinfo {pages} {8171} (\bibinfo {year}
  {1997})}\BibitemShut {NoStop}%
\bibitem [{\citenamefont {Coey}\ \emph {et~al.}(1999)\citenamefont {Coey},
  \citenamefont {Viret},\ and\ \citenamefont {von Molnar}}]{CVV99}%
  \BibitemOpen
  \bibfield  {author} {\bibinfo {author} {\bibfnamefont {J.~M.~D.}\
  \bibnamefont {Coey}}, \bibinfo {author} {\bibfnamefont {M.}~\bibnamefont
  {Viret}}, \ and\ \bibinfo {author} {\bibfnamefont {S.}~\bibnamefont {von
  Molnar}},\ }\href {\doibase 10.1080/000187399243455} {\bibfield  {journal}
  {\bibinfo  {journal} {Advances in Physics}\ }\textbf {\bibinfo {volume}
  {48}},\ \bibinfo {pages} {167} (\bibinfo {year} {1999})}\BibitemShut
  {NoStop}%
\bibitem [{\citenamefont {Aruta}\ \emph {et~al.}(2006)\citenamefont {Aruta},
  \citenamefont {Angeloni}, \citenamefont {Balestrino}, \citenamefont {Boggio},
  \citenamefont {Medaglia}, \citenamefont {Tebano}, \citenamefont {Davidson},
  \citenamefont {Baldini}, \citenamefont {{Di Castro}}, \citenamefont
  {Postorino}, \citenamefont {Dore}, \citenamefont {Sidorenko}, \citenamefont
  {Allodi},\ and\ \citenamefont {{De Renzi}}}]{AAB06}%
  \BibitemOpen
  \bibfield  {author} {\bibinfo {author} {\bibfnamefont {C.}~\bibnamefont
  {Aruta}}, \bibinfo {author} {\bibfnamefont {M.}~\bibnamefont {Angeloni}},
  \bibinfo {author} {\bibfnamefont {G.}~\bibnamefont {Balestrino}}, \bibinfo
  {author} {\bibfnamefont {N.~G.}\ \bibnamefont {Boggio}}, \bibinfo {author}
  {\bibfnamefont {P.~G.}\ \bibnamefont {Medaglia}}, \bibinfo {author}
  {\bibfnamefont {A.}~\bibnamefont {Tebano}}, \bibinfo {author} {\bibfnamefont
  {B.}~\bibnamefont {Davidson}}, \bibinfo {author} {\bibfnamefont
  {M.}~\bibnamefont {Baldini}}, \bibinfo {author} {\bibfnamefont
  {D.}~\bibnamefont {{Di Castro}}}, \bibinfo {author} {\bibfnamefont
  {P.}~\bibnamefont {Postorino}}, \bibinfo {author} {\bibfnamefont
  {P.}~\bibnamefont {Dore}}, \bibinfo {author} {\bibfnamefont {A.}~\bibnamefont
  {Sidorenko}}, \bibinfo {author} {\bibfnamefont {G.}~\bibnamefont {Allodi}}, \
  and\ \bibinfo {author} {\bibfnamefont {R.}~\bibnamefont {{De Renzi}}},\
  }\href {\doibase 10.1063/1.2217983} {\bibfield  {journal} {\bibinfo
  {journal} {Journal of Applied Physics}\ }\textbf {\bibinfo {volume} {100}},\
  \bibinfo {pages} {23910} (\bibinfo {year} {2006})}\BibitemShut {NoStop}%
\bibitem [{\citenamefont {Choi}\ \emph {et~al.}(2009)\citenamefont {Choi},
  \citenamefont {Marton}, \citenamefont {Jang}, \citenamefont {Moon},
  \citenamefont {Jeon}, \citenamefont {Shin}, \citenamefont {Seo},
  \citenamefont {Noh}, \citenamefont {Myung-Whun}, \citenamefont {Lee},\ and\
  \citenamefont {Lee}}]{CMJ09}%
  \BibitemOpen
  \bibfield  {author} {\bibinfo {author} {\bibfnamefont {W.~S.}\ \bibnamefont
  {Choi}}, \bibinfo {author} {\bibfnamefont {Z.}~\bibnamefont {Marton}},
  \bibinfo {author} {\bibfnamefont {S.~Y.}\ \bibnamefont {Jang}}, \bibinfo
  {author} {\bibfnamefont {S.~J.}\ \bibnamefont {Moon}}, \bibinfo {author}
  {\bibfnamefont {B.~C.}\ \bibnamefont {Jeon}}, \bibinfo {author}
  {\bibfnamefont {J.~H.}\ \bibnamefont {Shin}}, \bibinfo {author}
  {\bibfnamefont {S.~S.~a.}\ \bibnamefont {Seo}}, \bibinfo {author}
  {\bibfnamefont {T.~W.}\ \bibnamefont {Noh}}, \bibinfo {author} {\bibfnamefont
  {K.}~\bibnamefont {Myung-Whun}}, \bibinfo {author} {\bibfnamefont {H.~N.}\
  \bibnamefont {Lee}}, \ and\ \bibinfo {author} {\bibfnamefont {Y.~S.}\
  \bibnamefont {Lee}},\ }\href {\doibase 10.1088/0022-3727/42/16/165401}
  {\bibfield  {journal} {\bibinfo  {journal} {Journal of Physics D: Applied
  Physics}\ }\textbf {\bibinfo {volume} {42}},\ \bibinfo {pages} {165401}
  (\bibinfo {year} {2009})}\BibitemShut {NoStop}%
\bibitem [{\citenamefont {Marton}\ \emph {et~al.}(2010)\citenamefont {Marton},
  \citenamefont {{A. Seo}}, \citenamefont {Egami},\ and\ \citenamefont
  {Lee}}]{MSE10}%
  \BibitemOpen
  \bibfield  {author} {\bibinfo {author} {\bibfnamefont {Z.}~\bibnamefont
  {Marton}}, \bibinfo {author} {\bibfnamefont {S.~S.}\ \bibnamefont {{A.
  Seo}}}, \bibinfo {author} {\bibfnamefont {T.}~\bibnamefont {Egami}}, \ and\
  \bibinfo {author} {\bibfnamefont {H.~N.}\ \bibnamefont {Lee}},\ }\href
  {\doibase 10.1016/j.jcrysgro.2010.07.013} {\bibfield  {journal} {\bibinfo
  {journal} {Journal of Crystal Growth}\ }\textbf {\bibinfo {volume} {312}},\
  \bibinfo {pages} {2923} (\bibinfo {year} {2010})}\BibitemShut {NoStop}%
\bibitem [{\citenamefont {Gibert}\ \emph {et~al.}(2012)\citenamefont {Gibert},
  \citenamefont {Zubko}, \citenamefont {Scherwitzl}, \citenamefont
  {I\~{n}iguez},\ and\ \citenamefont {Triscone}}]{GZS12}%
  \BibitemOpen
  \bibfield  {author} {\bibinfo {author} {\bibfnamefont {M.}~\bibnamefont
  {Gibert}}, \bibinfo {author} {\bibfnamefont {P.}~\bibnamefont {Zubko}},
  \bibinfo {author} {\bibfnamefont {R.}~\bibnamefont {Scherwitzl}}, \bibinfo
  {author} {\bibfnamefont {J.}~\bibnamefont {I\~{n}iguez}}, \ and\ \bibinfo
  {author} {\bibfnamefont {J.-m.}\ \bibnamefont {Triscone}},\ }\href {\doibase
  10.1038/nmat3224} {\bibfield  {journal} {\bibinfo  {journal} {Nature
  Materials}\ }\textbf {\bibinfo {volume} {11}},\ \bibinfo {pages} {195}
  (\bibinfo {year} {2012})}\BibitemShut {NoStop}%
\bibitem [{\citenamefont {Sanchez}\ \emph {et~al.}(1996)\citenamefont
  {Sanchez}, \citenamefont {Causa}, \citenamefont {Caneiro}, \citenamefont
  {Butera}, \citenamefont {Vallet-Regi}, \citenamefont {Sayagues},
  \citenamefont {Gonzalez-Calbet}, \citenamefont {Garcia-Sanz},\ and\
  \citenamefont {Rivas}}]{SCC96}%
  \BibitemOpen
  \bibfield  {author} {\bibinfo {author} {\bibfnamefont {R.~D.}\ \bibnamefont
  {Sanchez}}, \bibinfo {author} {\bibfnamefont {M.~T.}\ \bibnamefont {Causa}},
  \bibinfo {author} {\bibfnamefont {A.}~\bibnamefont {Caneiro}}, \bibinfo
  {author} {\bibfnamefont {A.}~\bibnamefont {Butera}}, \bibinfo {author}
  {\bibfnamefont {M.}~\bibnamefont {Vallet-Regi}}, \bibinfo {author}
  {\bibfnamefont {M.~J.}\ \bibnamefont {Sayagues}}, \bibinfo {author}
  {\bibfnamefont {J.}~\bibnamefont {Gonzalez-Calbet}}, \bibinfo {author}
  {\bibfnamefont {F.}~\bibnamefont {Garcia-Sanz}}, \ and\ \bibinfo {author}
  {\bibfnamefont {J.}~\bibnamefont {Rivas}},\ }\href {\doibase
  10.1103/PhysRevB.54.16574} {\bibfield  {journal} {\bibinfo  {journal}
  {Physical Review B}\ }\textbf {\bibinfo {volume} {54}},\ \bibinfo {pages}
  {16574} (\bibinfo {year} {1996})}\BibitemShut {NoStop}%
\bibitem [{\citenamefont {Lekshmi}\ \emph {et~al.}(2005)\citenamefont
  {Lekshmi}, \citenamefont {Gayen}, \citenamefont {Sarma}, \citenamefont
  {Hegde}, \citenamefont {Chockalingam},\ and\ \citenamefont
  {Chandrasekhar}}]{LGS05}%
  \BibitemOpen
  \bibfield  {author} {\bibinfo {author} {\bibfnamefont {I.~C.}\ \bibnamefont
  {Lekshmi}}, \bibinfo {author} {\bibfnamefont {A.}~\bibnamefont {Gayen}},
  \bibinfo {author} {\bibfnamefont {D.~D.}\ \bibnamefont {Sarma}}, \bibinfo
  {author} {\bibfnamefont {M.~S.}\ \bibnamefont {Hegde}}, \bibinfo {author}
  {\bibfnamefont {S.~P.}\ \bibnamefont {Chockalingam}}, \ and\ \bibinfo
  {author} {\bibfnamefont {N.}~\bibnamefont {Chandrasekhar}},\ }\href {\doibase
  10.1063/1.2128046} {\bibfield  {journal} {\bibinfo  {journal} {Journal of
  Applied Physics}\ }\textbf {\bibinfo {volume} {98}},\ \bibinfo {pages}
  {093527} (\bibinfo {year} {2005})}\BibitemShut {NoStop}%
\bibitem [{\citenamefont {Xu}\ \emph {et~al.}(1993)\citenamefont {Xu},
  \citenamefont {Peng}, \citenamefont {Li}, \citenamefont {Ju},\ and\
  \citenamefont {Greene}}]{XPL93}%
  \BibitemOpen
  \bibfield  {author} {\bibinfo {author} {\bibfnamefont {X.~Q.}\ \bibnamefont
  {Xu}}, \bibinfo {author} {\bibfnamefont {J.~L.}\ \bibnamefont {Peng}},
  \bibinfo {author} {\bibfnamefont {Z.~Y.}\ \bibnamefont {Li}}, \bibinfo
  {author} {\bibfnamefont {H.~L.}\ \bibnamefont {Ju}}, \ and\ \bibinfo {author}
  {\bibfnamefont {R.~L.}\ \bibnamefont {Greene}},\ }\href {\doibase
  10.1103/PhysRevB.48.1112} {\bibfield  {journal} {\bibinfo  {journal}
  {Physical Review B}\ }\textbf {\bibinfo {volume} {48}},\ \bibinfo {pages}
  {1112} (\bibinfo {year} {1993})}\BibitemShut {NoStop}%
\bibitem [{\citenamefont {Gayathri}\ \emph {et~al.}(1998)\citenamefont
  {Gayathri}, \citenamefont {Raychaudhuri}, \citenamefont {Xu}, \citenamefont
  {Peng},\ and\ \citenamefont {Greene}}]{GRX98}%
  \BibitemOpen
  \bibfield  {author} {\bibinfo {author} {\bibfnamefont {N.}~\bibnamefont
  {Gayathri}}, \bibinfo {author} {\bibfnamefont {A.~K.}\ \bibnamefont
  {Raychaudhuri}}, \bibinfo {author} {\bibfnamefont {X.~Q.}\ \bibnamefont
  {Xu}}, \bibinfo {author} {\bibfnamefont {J.~L.}\ \bibnamefont {Peng}}, \ and\
  \bibinfo {author} {\bibfnamefont {R.~L.}\ \bibnamefont {Greene}},\ }\href
  {\doibase 10.1088/0953-8984/10/6/015} {\bibfield  {journal} {\bibinfo
  {journal} {Journal of Physics: Condensed Matter}\ }\textbf {\bibinfo {volume}
  {10}},\ \bibinfo {pages} {1323} (\bibinfo {year} {1998})}\BibitemShut
  {NoStop}%
\bibitem [{\citenamefont {May}\ \emph {et~al.}(2009)\citenamefont {May},
  \citenamefont {Santos},\ and\ \citenamefont {Bhattacharya}}]{MSB09}%
  \BibitemOpen
  \bibfield  {author} {\bibinfo {author} {\bibfnamefont {S.~J.}\ \bibnamefont
  {May}}, \bibinfo {author} {\bibfnamefont {T.~S.}\ \bibnamefont {Santos}}, \
  and\ \bibinfo {author} {\bibfnamefont {A.}~\bibnamefont {Bhattacharya}},\
  }\href {\doibase 10.1103/PhysRevB.79.115127} {\bibfield  {journal} {\bibinfo
  {journal} {Physical Review B}\ }\textbf {\bibinfo {volume} {79}},\ \bibinfo
  {pages} {115127} (\bibinfo {year} {2009})}\BibitemShut {NoStop}%
\bibitem [{\citenamefont {Santos}\ \emph {et~al.}(2009)\citenamefont {Santos},
  \citenamefont {May}, \citenamefont {Robertson},\ and\ \citenamefont
  {Bhattacharya}}]{SMR09}%
  \BibitemOpen
  \bibfield  {author} {\bibinfo {author} {\bibfnamefont {T.~S.}\ \bibnamefont
  {Santos}}, \bibinfo {author} {\bibfnamefont {S.~J.}\ \bibnamefont {May}},
  \bibinfo {author} {\bibfnamefont {J.~L.}\ \bibnamefont {Robertson}}, \ and\
  \bibinfo {author} {\bibfnamefont {A.}~\bibnamefont {Bhattacharya}},\ }\href
  {\doibase 10.1103/PhysRevB.80.155114} {\bibfield  {journal} {\bibinfo
  {journal} {Physical Review B}\ }\textbf {\bibinfo {volume} {80}},\ \bibinfo
  {pages} {155114} (\bibinfo {year} {2009})}\BibitemShut {NoStop}%
\bibitem [{Note1()}]{Note1}%
  \BibitemOpen
  \bibinfo {note} {The out-of-plane lattice constants for 80 unit cell thick
  films of LNO and LMO grown on (001) SrTiO$_3$ substrates were measured and
  found to be 0.394 nm and 0.381 nm, respectively.}\BibitemShut {Stop}%
\bibitem [{\citenamefont {Abbate}\ \emph {et~al.}(1992)\citenamefont {Abbate},
  \citenamefont {de~Groot}, \citenamefont {Fuggle}, \citenamefont {Fujimori},
  \citenamefont {Strebel}, \citenamefont {Lopez}, \citenamefont {Domke},
  \citenamefont {Kaindl}, \citenamefont {Sawatzky}, \citenamefont {Takano},
  \citenamefont {Takeda}, \citenamefont {Eisaki},\ and\ \citenamefont
  {Uchida}}]{ADF92}%
  \BibitemOpen
  \bibfield  {author} {\bibinfo {author} {\bibfnamefont {M.}~\bibnamefont
  {Abbate}}, \bibinfo {author} {\bibfnamefont {F.~M.~F.}\ \bibnamefont
  {de~Groot}}, \bibinfo {author} {\bibfnamefont {J.~C.}\ \bibnamefont
  {Fuggle}}, \bibinfo {author} {\bibfnamefont {A.}~\bibnamefont {Fujimori}},
  \bibinfo {author} {\bibfnamefont {O.}~\bibnamefont {Strebel}}, \bibinfo
  {author} {\bibfnamefont {F.}~\bibnamefont {Lopez}}, \bibinfo {author}
  {\bibfnamefont {M.}~\bibnamefont {Domke}}, \bibinfo {author} {\bibfnamefont
  {G.}~\bibnamefont {Kaindl}}, \bibinfo {author} {\bibfnamefont {G.~A.}\
  \bibnamefont {Sawatzky}}, \bibinfo {author} {\bibfnamefont {M.}~\bibnamefont
  {Takano}}, \bibinfo {author} {\bibfnamefont {Y.}~\bibnamefont {Takeda}},
  \bibinfo {author} {\bibfnamefont {H.}~\bibnamefont {Eisaki}}, \ and\ \bibinfo
  {author} {\bibfnamefont {S.}~\bibnamefont {Uchida}},\ }\href {\doibase
  10.1103/PhysRevB.46.4511} {\bibfield  {journal} {\bibinfo  {journal}
  {Physical Review B}\ }\textbf {\bibinfo {volume} {46}},\ \bibinfo {pages}
  {4511} (\bibinfo {year} {1992})}\BibitemShut {NoStop}%
\bibitem [{\citenamefont {Medarde}(1997)}]{M97}%
  \BibitemOpen
  \bibfield  {author} {\bibinfo {author} {\bibfnamefont {M.~L.}\ \bibnamefont
  {Medarde}},\ }\href {\doibase 10.1088/0953-8984/9/8/003} {\bibfield
  {journal} {\bibinfo  {journal} {Journal of Physics: Condensed Matter}\
  }\textbf {\bibinfo {volume} {9}},\ \bibinfo {pages} {1679} (\bibinfo {year}
  {1997})}\BibitemShut {NoStop}%
\bibitem [{\citenamefont {Liu}\ \emph {et~al.}(2011)\citenamefont {Liu},
  \citenamefont {Okamoto}, \citenamefont {van Veenendaal}, \citenamefont
  {Kareev}, \citenamefont {Gray}, \citenamefont {Ryan}, \citenamefont
  {Freeland},\ and\ \citenamefont {Chakhalian}}]{LOV11}%
  \BibitemOpen
  \bibfield  {author} {\bibinfo {author} {\bibfnamefont {J.}~\bibnamefont
  {Liu}}, \bibinfo {author} {\bibfnamefont {S.}~\bibnamefont {Okamoto}},
  \bibinfo {author} {\bibfnamefont {M.}~\bibnamefont {van Veenendaal}},
  \bibinfo {author} {\bibfnamefont {M.}~\bibnamefont {Kareev}}, \bibinfo
  {author} {\bibfnamefont {B.}~\bibnamefont {Gray}}, \bibinfo {author}
  {\bibfnamefont {P.}~\bibnamefont {Ryan}}, \bibinfo {author} {\bibfnamefont
  {J.~W.}\ \bibnamefont {Freeland}}, \ and\ \bibinfo {author} {\bibfnamefont
  {J.}~\bibnamefont {Chakhalian}},\ }\href {\doibase
  10.1103/PhysRevB.83.161102} {\bibfield  {journal} {\bibinfo  {journal}
  {Physical Review B}\ }\textbf {\bibinfo {volume} {83}},\ \bibinfo {pages}
  {161102} (\bibinfo {year} {2011})}\BibitemShut {NoStop}%
\bibitem [{\citenamefont {Freeland}\ \emph {et~al.}(2011)\citenamefont
  {Freeland}, \citenamefont {Liu}, \citenamefont {Kareev}, \citenamefont
  {Gray}, \citenamefont {Kim}, \citenamefont {Ryan}, \citenamefont
  {Pentcheva},\ and\ \citenamefont {Chakhalian}}]{FLK11}%
  \BibitemOpen
  \bibfield  {author} {\bibinfo {author} {\bibfnamefont {J.~W.}\ \bibnamefont
  {Freeland}}, \bibinfo {author} {\bibfnamefont {J.}~\bibnamefont {Liu}},
  \bibinfo {author} {\bibfnamefont {M.}~\bibnamefont {Kareev}}, \bibinfo
  {author} {\bibfnamefont {B.}~\bibnamefont {Gray}}, \bibinfo {author}
  {\bibfnamefont {J.~W.}\ \bibnamefont {Kim}}, \bibinfo {author} {\bibfnamefont
  {P.}~\bibnamefont {Ryan}}, \bibinfo {author} {\bibfnamefont {R.}~\bibnamefont
  {Pentcheva}}, \ and\ \bibinfo {author} {\bibfnamefont {J.}~\bibnamefont
  {Chakhalian}},\ }\href {\doibase 10.1209/0295-5075/96/57004} {\bibfield
  {journal} {\bibinfo  {journal} {Europhysics Letters}\ }\textbf {\bibinfo
  {volume} {96}},\ \bibinfo {pages} {57004} (\bibinfo {year}
  {2011})}\BibitemShut {NoStop}%
\bibitem [{\citenamefont {Guo}\ \emph {et~al.}(2009)\citenamefont {Guo},
  \citenamefont {Gupta}, \citenamefont {Varela}, \citenamefont {Pennycook},\
  and\ \citenamefont {Zhang}}]{GGV09}%
  \BibitemOpen
  \bibfield  {author} {\bibinfo {author} {\bibfnamefont {H.}~\bibnamefont
  {Guo}}, \bibinfo {author} {\bibfnamefont {A.}~\bibnamefont {Gupta}}, \bibinfo
  {author} {\bibfnamefont {M.}~\bibnamefont {Varela}}, \bibinfo {author}
  {\bibfnamefont {S.~J.}\ \bibnamefont {Pennycook}}, \ and\ \bibinfo {author}
  {\bibfnamefont {J.}~\bibnamefont {Zhang}},\ }\href {\doibase
  10.1103/PhysRevB.79.172402} {\bibfield  {journal} {\bibinfo  {journal}
  {Physical Review B}\ }\textbf {\bibinfo {volume} {79}},\ \bibinfo {pages}
  {172402} (\bibinfo {year} {2009})}\BibitemShut {NoStop}%
\bibitem [{\citenamefont {{Rojas S\'{a}nchez}}\ \emph
  {et~al.}(2012)\citenamefont {{Rojas S\'{a}nchez}}, \citenamefont
  {Nelson-Cheeseman}, \citenamefont {Granada}, \citenamefont {Arenholz},\ and\
  \citenamefont {Steren}}]{SNG12}%
  \BibitemOpen
  \bibfield  {author} {\bibinfo {author} {\bibfnamefont {J.~C.}\ \bibnamefont
  {{Rojas S\'{a}nchez}}}, \bibinfo {author} {\bibfnamefont {B.}~\bibnamefont
  {Nelson-Cheeseman}}, \bibinfo {author} {\bibfnamefont {M.}~\bibnamefont
  {Granada}}, \bibinfo {author} {\bibfnamefont {E.}~\bibnamefont {Arenholz}}, \
  and\ \bibinfo {author} {\bibfnamefont {L.~B.}\ \bibnamefont {Steren}},\
  }\href {\doibase 10.1103/PhysRevB.85.094427} {\bibfield  {journal} {\bibinfo
  {journal} {Physical Review B}\ }\textbf {\bibinfo {volume} {85}},\ \bibinfo
  {pages} {94427} (\bibinfo {year} {2012})}\BibitemShut {NoStop}%
\bibitem [{Note2()}]{Note2}%
  \BibitemOpen
  \bibinfo {note} {To quantify the magnetization of the individual cations, we
  compare the amplitude of the XMCD spectra with calibrated reference samples
  where the amplitude of the XMCD signal and magnetization are
  known.}\BibitemShut {Stop}%
\bibitem [{\citenamefont {Neumeier}\ and\ \citenamefont {Cohn}(2000)}]{NC00}%
  \BibitemOpen
  \bibfield  {author} {\bibinfo {author} {\bibfnamefont {J.~J.}\ \bibnamefont
  {Neumeier}}\ and\ \bibinfo {author} {\bibfnamefont {J.~L.}\ \bibnamefont
  {Cohn}},\ }\href {\doibase 10.1103/PhysRevB.61.14319} {\bibfield  {journal}
  {\bibinfo  {journal} {Physical Review B}\ }\textbf {\bibinfo {volume} {61}},\
  \bibinfo {pages} {14319} (\bibinfo {year} {2000})}\BibitemShut {NoStop}%
\bibitem [{\citenamefont {Tsukahara}\ \emph {et~al.}(2010)\citenamefont
  {Tsukahara}, \citenamefont {Ishibashi},\ and\ \citenamefont
  {Terakura}}]{TIT10}%
  \BibitemOpen
  \bibfield  {author} {\bibinfo {author} {\bibfnamefont {H.}~\bibnamefont
  {Tsukahara}}, \bibinfo {author} {\bibfnamefont {S.}~\bibnamefont
  {Ishibashi}}, \ and\ \bibinfo {author} {\bibfnamefont {K.}~\bibnamefont
  {Terakura}},\ }\href {\doibase 10.1103/PhysRevB.81.214108} {\bibfield
  {journal} {\bibinfo  {journal} {Physical Review B}\ }\textbf {\bibinfo
  {volume} {81}},\ \bibinfo {pages} {214108} (\bibinfo {year}
  {2010})}\BibitemShut {NoStop}%
\bibitem [{Note3()}]{Note3}%
  \BibitemOpen
  \bibinfo {note} {The diamagnetic contribution due to the SrTiO$_3$ substrate
  was measured and subtracted from the total magnetization measured by SQUID
  magnetometry.}\BibitemShut {Stop}%
\bibitem [{\citenamefont {Shklovskii}\ and\ \citenamefont
  {Efros}(1984)}]{SE84}%
  \BibitemOpen
  \bibfield  {author} {\bibinfo {author} {\bibfnamefont {B.~I.}\ \bibnamefont
  {Shklovskii}}\ and\ \bibinfo {author} {\bibfnamefont {A.~L.}\ \bibnamefont
  {Efros}},\ }\href@noop {} {\emph {\bibinfo {title} {{Electronic Properties of
  Doped Semiconductors}}}}\ (\bibinfo  {publisher} {Springer-Verlag},\ \bibinfo
  {address} {New York},\ \bibinfo {year} {1984})\ p.\ \bibinfo {pages}
  {362}\BibitemShut {NoStop}%
\bibitem [{\citenamefont {Scherwitzl}\ \emph {et~al.}(2011)\citenamefont
  {Scherwitzl}, \citenamefont {Gariglio}, \citenamefont {Gabay}, \citenamefont
  {Zubko}, \citenamefont {Gibert},\ and\ \citenamefont {Triscone}}]{SGG11}%
  \BibitemOpen
  \bibfield  {author} {\bibinfo {author} {\bibfnamefont {R.}~\bibnamefont
  {Scherwitzl}}, \bibinfo {author} {\bibfnamefont {S.}~\bibnamefont
  {Gariglio}}, \bibinfo {author} {\bibfnamefont {M.}~\bibnamefont {Gabay}},
  \bibinfo {author} {\bibfnamefont {P.}~\bibnamefont {Zubko}}, \bibinfo
  {author} {\bibfnamefont {M.}~\bibnamefont {Gibert}}, \ and\ \bibinfo {author}
  {\bibfnamefont {J.-M.}\ \bibnamefont {Triscone}},\ }\href {\doibase
  10.1103/PhysRevLett.106.246403} {\bibfield  {journal} {\bibinfo  {journal}
  {Physical Review Letters}\ }\textbf {\bibinfo {volume} {106}},\ \bibinfo
  {pages} {246403} (\bibinfo {year} {2011})}\BibitemShut {NoStop}%
\bibitem [{\citenamefont {Rajeev}\ \emph {et~al.}(1991)\citenamefont {Rajeev},
  \citenamefont {Shivashankar},\ and\ \citenamefont {Raychaudhuri}}]{RSR91}%
  \BibitemOpen
  \bibfield  {author} {\bibinfo {author} {\bibfnamefont {K.~P.}\ \bibnamefont
  {Rajeev}}, \bibinfo {author} {\bibfnamefont {G.~V.}\ \bibnamefont
  {Shivashankar}}, \ and\ \bibinfo {author} {\bibfnamefont {A.~K.}\
  \bibnamefont {Raychaudhuri}},\ }\href {\doibase 10.1016/0038-1098(91)90915-I}
  {\bibfield  {journal} {\bibinfo  {journal} {Solid State Communications}\
  }\textbf {\bibinfo {volume} {79}},\ \bibinfo {pages} {591} (\bibinfo {year}
  {1991})}\BibitemShut {NoStop}%
\bibitem [{\citenamefont {Viret}\ \emph {et~al.}(1997)\citenamefont {Viret},
  \citenamefont {Ranno},\ and\ \citenamefont {Coey}}]{VRC97b}%
  \BibitemOpen
  \bibfield  {author} {\bibinfo {author} {\bibfnamefont {M.}~\bibnamefont
  {Viret}}, \bibinfo {author} {\bibfnamefont {L.}~\bibnamefont {Ranno}}, \ and\
  \bibinfo {author} {\bibfnamefont {J.~M.~D.}\ \bibnamefont {Coey}},\ }\href
  {\doibase 10.1103/PhysRevB.55.8067} {\bibfield  {journal} {\bibinfo
  {journal} {Physical Review B}\ }\textbf {\bibinfo {volume} {55}},\ \bibinfo
  {pages} {8067} (\bibinfo {year} {1997})}\BibitemShut {NoStop}%
\bibitem [{\citenamefont {Moreo}\ \emph {et~al.}(1999)\citenamefont {Moreo},
  \citenamefont {Yunoki},\ and\ \citenamefont {Dagotto}}]{MYD99}%
  \BibitemOpen
  \bibfield  {author} {\bibinfo {author} {\bibfnamefont {A.}~\bibnamefont
  {Moreo}}, \bibinfo {author} {\bibfnamefont {S.}~\bibnamefont {Yunoki}}, \
  and\ \bibinfo {author} {\bibfnamefont {E.}~\bibnamefont {Dagotto}},\ }\href
  {\doibase 10.1103/PhysRevLett.83.2773} {\bibfield  {journal} {\bibinfo
  {journal} {Physical Review Letters}\ }\textbf {\bibinfo {volume} {83}},\
  \bibinfo {pages} {2773} (\bibinfo {year} {1999})}\BibitemShut {NoStop}%
\bibitem [{\citenamefont {Boris}\ \emph {et~al.}(2011)\citenamefont {Boris},
  \citenamefont {Matiks}, \citenamefont {Benckiser}, \citenamefont {Frano},
  \citenamefont {Popovich}, \citenamefont {Hinkov}, \citenamefont {Wochner},
  \citenamefont {Castro-Colin}, \citenamefont {Detemple}, \citenamefont
  {Malik}, \citenamefont {Bernhard}, \citenamefont {Prokscha}, \citenamefont
  {Suter}, \citenamefont {Salman}, \citenamefont {Morenzoni}, \citenamefont
  {Cristiani}, \citenamefont {Habermeier},\ and\ \citenamefont
  {Keimer}}]{BMB11}%
  \BibitemOpen
  \bibfield  {author} {\bibinfo {author} {\bibfnamefont {A.~V.}\ \bibnamefont
  {Boris}}, \bibinfo {author} {\bibfnamefont {Y.}~\bibnamefont {Matiks}},
  \bibinfo {author} {\bibfnamefont {E.}~\bibnamefont {Benckiser}}, \bibinfo
  {author} {\bibfnamefont {A.}~\bibnamefont {Frano}}, \bibinfo {author}
  {\bibfnamefont {P.}~\bibnamefont {Popovich}}, \bibinfo {author}
  {\bibfnamefont {V.}~\bibnamefont {Hinkov}}, \bibinfo {author} {\bibfnamefont
  {P.}~\bibnamefont {Wochner}}, \bibinfo {author} {\bibfnamefont
  {M.}~\bibnamefont {Castro-Colin}}, \bibinfo {author} {\bibfnamefont
  {E.}~\bibnamefont {Detemple}}, \bibinfo {author} {\bibfnamefont {V.~K.}\
  \bibnamefont {Malik}}, \bibinfo {author} {\bibfnamefont {C.}~\bibnamefont
  {Bernhard}}, \bibinfo {author} {\bibfnamefont {T.}~\bibnamefont {Prokscha}},
  \bibinfo {author} {\bibfnamefont {A.}~\bibnamefont {Suter}}, \bibinfo
  {author} {\bibfnamefont {Z.}~\bibnamefont {Salman}}, \bibinfo {author}
  {\bibfnamefont {E.}~\bibnamefont {Morenzoni}}, \bibinfo {author}
  {\bibfnamefont {G.}~\bibnamefont {Cristiani}}, \bibinfo {author}
  {\bibfnamefont {H.-U.}\ \bibnamefont {Habermeier}}, \ and\ \bibinfo {author}
  {\bibfnamefont {B.}~\bibnamefont {Keimer}},\ }\href {\doibase
  10.1126/science.1202647} {\bibfield  {journal} {\bibinfo  {journal}
  {Science}\ }\textbf {\bibinfo {volume} {332}},\ \bibinfo {pages} {937}
  (\bibinfo {year} {2011})}\BibitemShut {NoStop}%
\bibitem [{\citenamefont {Son}\ \emph {et~al.}(2010{\natexlab{a}})\citenamefont
  {Son}, \citenamefont {LeBeau}, \citenamefont {Allen},\ and\ \citenamefont
  {Stemmer}}]{SLA10}%
  \BibitemOpen
  \bibfield  {author} {\bibinfo {author} {\bibfnamefont {J.}~\bibnamefont
  {Son}}, \bibinfo {author} {\bibfnamefont {J.~M.}\ \bibnamefont {LeBeau}},
  \bibinfo {author} {\bibfnamefont {S.~J.}\ \bibnamefont {Allen}}, \ and\
  \bibinfo {author} {\bibfnamefont {S.}~\bibnamefont {Stemmer}},\ }\href
  {\doibase 10.1063/1.3511738} {\bibfield  {journal} {\bibinfo  {journal}
  {Applied Physics Letters}\ }\textbf {\bibinfo {volume} {97}},\ \bibinfo
  {pages} {202109} (\bibinfo {year} {2010}{\natexlab{a}})}\BibitemShut
  {NoStop}%
\bibitem [{\citenamefont {Son}\ \emph {et~al.}(2010{\natexlab{b}})\citenamefont
  {Son}, \citenamefont {Moetakef}, \citenamefont {LeBeau}, \citenamefont
  {Ouellette}, \citenamefont {Balents}, \citenamefont {Allen},\ and\
  \citenamefont {Stemmer}}]{SML10}%
  \BibitemOpen
  \bibfield  {author} {\bibinfo {author} {\bibfnamefont {J.}~\bibnamefont
  {Son}}, \bibinfo {author} {\bibfnamefont {P.}~\bibnamefont {Moetakef}},
  \bibinfo {author} {\bibfnamefont {J.~M.}\ \bibnamefont {LeBeau}}, \bibinfo
  {author} {\bibfnamefont {D.}~\bibnamefont {Ouellette}}, \bibinfo {author}
  {\bibfnamefont {L.}~\bibnamefont {Balents}}, \bibinfo {author} {\bibfnamefont
  {S.~J.}\ \bibnamefont {Allen}}, \ and\ \bibinfo {author} {\bibfnamefont
  {S.}~\bibnamefont {Stemmer}},\ }\href {\doibase 10.1063/1.3309713} {\bibfield
   {journal} {\bibinfo  {journal} {Applied Physics Letters}\ }\textbf {\bibinfo
  {volume} {96}},\ \bibinfo {pages} {62114} (\bibinfo {year}
  {2010}{\natexlab{b}})}\BibitemShut {NoStop}%
\bibitem [{\citenamefont {Shinomori}\ \emph {et~al.}(2002)\citenamefont
  {Shinomori}, \citenamefont {Okimoto}, \citenamefont {Kawasaki},\ and\
  \citenamefont {Tokura}}]{SOK02}%
  \BibitemOpen
  \bibfield  {author} {\bibinfo {author} {\bibfnamefont {S.}~\bibnamefont
  {Shinomori}}, \bibinfo {author} {\bibfnamefont {Y.}~\bibnamefont {Okimoto}},
  \bibinfo {author} {\bibfnamefont {M.}~\bibnamefont {Kawasaki}}, \ and\
  \bibinfo {author} {\bibfnamefont {Y.}~\bibnamefont {Tokura}},\ }\href
  {\doibase 10.1143/JPSJ.71.705} {\bibfield  {journal} {\bibinfo  {journal}
  {Journal of the Physical Society of Japan}\ }\textbf {\bibinfo {volume}
  {71}},\ \bibinfo {pages} {705} (\bibinfo {year} {2002})}\BibitemShut
  {NoStop}%
\bibitem [{\citenamefont {Hamada}(1993)}]{H93}%
  \BibitemOpen
  \bibfield  {author} {\bibinfo {author} {\bibfnamefont {N.}~\bibnamefont
  {Hamada}},\ }\href {\doibase 10.1016/0022-3697(93)90159-O} {\bibfield
  {journal} {\bibinfo  {journal} {Journal of Physics and Chemistry of Solids}\
  }\textbf {\bibinfo {volume} {54}},\ \bibinfo {pages} {1157} (\bibinfo {year}
  {1993})}\BibitemShut {NoStop}%
\bibitem [{\citenamefont {Carter}\ \emph {et~al.}(1993)\citenamefont {Carter},
  \citenamefont {Rosenbaum}, \citenamefont {Metcalf}, \citenamefont {Honig},\
  and\ \citenamefont {Spalek}}]{CRM93}%
  \BibitemOpen
  \bibfield  {author} {\bibinfo {author} {\bibfnamefont {S.~A.}\ \bibnamefont
  {Carter}}, \bibinfo {author} {\bibfnamefont {T.~F.}\ \bibnamefont
  {Rosenbaum}}, \bibinfo {author} {\bibfnamefont {P.}~\bibnamefont {Metcalf}},
  \bibinfo {author} {\bibfnamefont {J.~M.}\ \bibnamefont {Honig}}, \ and\
  \bibinfo {author} {\bibfnamefont {J.}~\bibnamefont {Spalek}},\ }\href
  {\doibase 10.1103/PhysRevB.48.16841} {\bibfield  {journal} {\bibinfo
  {journal} {Physical Review B}\ }\textbf {\bibinfo {volume} {48}},\ \bibinfo
  {pages} {16841} (\bibinfo {year} {1993})}\BibitemShut {NoStop}%
\bibitem [{\citenamefont {Miyasaka}\ \emph {et~al.}(2000)\citenamefont
  {Miyasaka}, \citenamefont {Takagi}, \citenamefont {Sekine}, \citenamefont
  {Takahashi}, \citenamefont {M\^{o}ri},\ and\ \citenamefont {Cava}}]{MTS00}%
  \BibitemOpen
  \bibfield  {author} {\bibinfo {author} {\bibfnamefont {S.}~\bibnamefont
  {Miyasaka}}, \bibinfo {author} {\bibfnamefont {H.}~\bibnamefont {Takagi}},
  \bibinfo {author} {\bibfnamefont {Y.}~\bibnamefont {Sekine}}, \bibinfo
  {author} {\bibfnamefont {H.}~\bibnamefont {Takahashi}}, \bibinfo {author}
  {\bibfnamefont {N.}~\bibnamefont {M\^{o}ri}}, \ and\ \bibinfo {author}
  {\bibfnamefont {R.~J.}\ \bibnamefont {Cava}},\ }\href {\doibase
  10.1143/JPSJ.69.3166} {\bibfield  {journal} {\bibinfo  {journal} {Journal of
  the Physical Society of Japan}\ }\textbf {\bibinfo {volume} {69}},\ \bibinfo
  {pages} {3166} (\bibinfo {year} {2000})}\BibitemShut {NoStop}%
\bibitem [{\citenamefont {Nishikawa}\ \emph {et~al.}(1993)\citenamefont
  {Nishikawa}, \citenamefont {Takeda},\ and\ \citenamefont {Sato}}]{NTS93}%
  \BibitemOpen
  \bibfield  {author} {\bibinfo {author} {\bibfnamefont {T.}~\bibnamefont
  {Nishikawa}}, \bibinfo {author} {\bibfnamefont {J.}~\bibnamefont {Takeda}}, \
  and\ \bibinfo {author} {\bibfnamefont {M.}~\bibnamefont {Sato}},\ }\href
  {\doibase 10.1143/JPSJ.62.2568} {\bibfield  {journal} {\bibinfo  {journal}
  {Journal of the Physical Society of Japan}\ }\textbf {\bibinfo {volume}
  {62}},\ \bibinfo {pages} {2568} (\bibinfo {year} {1993})}\BibitemShut
  {NoStop}%
\bibitem [{\citenamefont {Ito}\ \emph {et~al.}(1993)\citenamefont {Ito},
  \citenamefont {Takenaka},\ and\ \citenamefont {Uchida}}]{IKU93}%
  \BibitemOpen
  \bibfield  {author} {\bibinfo {author} {\bibfnamefont {T.}~\bibnamefont
  {Ito}}, \bibinfo {author} {\bibfnamefont {K.}~\bibnamefont {Takenaka}}, \
  and\ \bibinfo {author} {\bibfnamefont {S.}~\bibnamefont {Uchida}},\ }\href
  {\doibase 10.1103/PhysRevLett.70.3995} {\bibfield  {journal} {\bibinfo
  {journal} {Physical Review Letters}\ }\textbf {\bibinfo {volume} {70}},\
  \bibinfo {pages} {3995} (\bibinfo {year} {1993})}\BibitemShut {NoStop}%
\bibitem [{\citenamefont {Gordon}\ \emph {et~al.}(2000)\citenamefont {Gordon},
  \citenamefont {Wagner}, \citenamefont {Das}, \citenamefont {Vanacken},
  \citenamefont {Moshchalkov}, \citenamefont {Bruynseraede}, \citenamefont
  {Schuddinck}, \citenamefont {{Van Tendeloo}}, \citenamefont {Ziese},\ and\
  \citenamefont {Borghs}}]{GWD00}%
  \BibitemOpen
  \bibfield  {author} {\bibinfo {author} {\bibfnamefont {I.}~\bibnamefont
  {Gordon}}, \bibinfo {author} {\bibfnamefont {P.}~\bibnamefont {Wagner}},
  \bibinfo {author} {\bibfnamefont {A.}~\bibnamefont {Das}}, \bibinfo {author}
  {\bibfnamefont {J.}~\bibnamefont {Vanacken}}, \bibinfo {author}
  {\bibfnamefont {V.~V.}\ \bibnamefont {Moshchalkov}}, \bibinfo {author}
  {\bibfnamefont {Y.}~\bibnamefont {Bruynseraede}}, \bibinfo {author}
  {\bibfnamefont {W.}~\bibnamefont {Schuddinck}}, \bibinfo {author}
  {\bibfnamefont {G.}~\bibnamefont {{Van Tendeloo}}}, \bibinfo {author}
  {\bibfnamefont {M.}~\bibnamefont {Ziese}}, \ and\ \bibinfo {author}
  {\bibfnamefont {G.}~\bibnamefont {Borghs}},\ }\href {\doibase
  10.1103/PhysRevB.62.11633} {\bibfield  {journal} {\bibinfo  {journal}
  {Physical Review B}\ }\textbf {\bibinfo {volume} {62}},\ \bibinfo {pages}
  {11633} (\bibinfo {year} {2000})}\BibitemShut {NoStop}%
\bibitem [{\citenamefont {Yang}\ \emph {et~al.}(2001)\citenamefont {Yang},
  \citenamefont {Wang},\ and\ \citenamefont {Horng}}]{YWH01}%
  \BibitemOpen
  \bibfield  {author} {\bibinfo {author} {\bibfnamefont {H.~C.}\ \bibnamefont
  {Yang}}, \bibinfo {author} {\bibfnamefont {L.~M.}\ \bibnamefont {Wang}}, \
  and\ \bibinfo {author} {\bibfnamefont {H.~E.}\ \bibnamefont {Horng}},\ }\href
  {\doibase 10.1103/PhysRevB.64.174415} {\bibfield  {journal} {\bibinfo
  {journal} {Physical Review B}\ }\textbf {\bibinfo {volume} {64}},\ \bibinfo
  {pages} {174415} (\bibinfo {year} {2001})}\BibitemShut {NoStop}%
\bibitem [{\citenamefont {Bebenin}\ \emph {et~al.}(2004)\citenamefont
  {Bebenin}, \citenamefont {Zainullina}, \citenamefont {Mashkautsan},
  \citenamefont {Ustinov},\ and\ \citenamefont {Mukovskii}}]{BZM04}%
  \BibitemOpen
  \bibfield  {author} {\bibinfo {author} {\bibfnamefont {N.~G.}\ \bibnamefont
  {Bebenin}}, \bibinfo {author} {\bibfnamefont {R.~I.}\ \bibnamefont
  {Zainullina}}, \bibinfo {author} {\bibfnamefont {V.~V.}\ \bibnamefont
  {Mashkautsan}}, \bibinfo {author} {\bibfnamefont {V.~V.}\ \bibnamefont
  {Ustinov}}, \ and\ \bibinfo {author} {\bibfnamefont {Y.~M.}\ \bibnamefont
  {Mukovskii}},\ }\href {\doibase 10.1103/PhysRevB.69.104434} {\bibfield
  {journal} {\bibinfo  {journal} {Physical Review B}\ }\textbf {\bibinfo
  {volume} {69}},\ \bibinfo {pages} {104434} (\bibinfo {year}
  {2004})}\BibitemShut {NoStop}%
\bibitem [{\citenamefont {Naftalis}\ \emph {et~al.}(2012)\citenamefont
  {Naftalis}, \citenamefont {Haham}, \citenamefont {Hoffman}, \citenamefont
  {Marshall}, \citenamefont {Ahn},\ and\ \citenamefont {Klein}}]{NHH12}%
  \BibitemOpen
  \bibfield  {author} {\bibinfo {author} {\bibfnamefont {N.}~\bibnamefont
  {Naftalis}}, \bibinfo {author} {\bibfnamefont {N.}~\bibnamefont {Haham}},
  \bibinfo {author} {\bibfnamefont {J.}~\bibnamefont {Hoffman}}, \bibinfo
  {author} {\bibfnamefont {M.~S.~J.}\ \bibnamefont {Marshall}}, \bibinfo
  {author} {\bibfnamefont {C.~H.}\ \bibnamefont {Ahn}}, \ and\ \bibinfo
  {author} {\bibfnamefont {L.}~\bibnamefont {Klein}},\ }\href {\doibase
  10.1103/PhysRevB.86.184402} {\bibfield  {journal} {\bibinfo  {journal}
  {Physical Review B}\ }\textbf {\bibinfo {volume} {86}},\ \bibinfo {pages}
  {184402} (\bibinfo {year} {2012})}\BibitemShut {NoStop}%
\bibitem [{\citenamefont {Nagaosa}\ \emph {et~al.}(2010)\citenamefont
  {Nagaosa}, \citenamefont {Sinova}, \citenamefont {Onoda}, \citenamefont
  {MacDonald},\ and\ \citenamefont {Ong}}]{NSO10}%
  \BibitemOpen
  \bibfield  {author} {\bibinfo {author} {\bibfnamefont {N.}~\bibnamefont
  {Nagaosa}}, \bibinfo {author} {\bibfnamefont {J.}~\bibnamefont {Sinova}},
  \bibinfo {author} {\bibfnamefont {S.}~\bibnamefont {Onoda}}, \bibinfo
  {author} {\bibfnamefont {A.~H.}\ \bibnamefont {MacDonald}}, \ and\ \bibinfo
  {author} {\bibfnamefont {N.~P.}\ \bibnamefont {Ong}},\ }\href {\doibase
  10.1103/RevModPhys.82.1539} {\bibfield  {journal} {\bibinfo  {journal}
  {Reviews of Modern Physics}\ }\textbf {\bibinfo {volume} {82}},\ \bibinfo
  {pages} {1539} (\bibinfo {year} {2010})}\BibitemShut {NoStop}%
\bibitem [{\citenamefont {Canedy}\ \emph {et~al.}(2000)\citenamefont {Canedy},
  \citenamefont {Li},\ and\ \citenamefont {Xiao}}]{CLX00}%
  \BibitemOpen
  \bibfield  {author} {\bibinfo {author} {\bibfnamefont {C.~L.}\ \bibnamefont
  {Canedy}}, \bibinfo {author} {\bibfnamefont {X.~W.}\ \bibnamefont {Li}}, \
  and\ \bibinfo {author} {\bibfnamefont {G.}~\bibnamefont {Xiao}},\ }\href
  {\doibase 10.1103/PhysRevB.62.508} {\bibfield  {journal} {\bibinfo  {journal}
  {Physical Review B}\ }\textbf {\bibinfo {volume} {62}},\ \bibinfo {pages}
  {508} (\bibinfo {year} {2000})}\BibitemShut {NoStop}%
\bibitem [{\citenamefont {Xiong}\ \emph {et~al.}(1992)\citenamefont {Xiong},
  \citenamefont {Xiao}, \citenamefont {Wang}, \citenamefont {Xiao},
  \citenamefont {Jiang},\ and\ \citenamefont {Chien}}]{XXW92}%
  \BibitemOpen
  \bibfield  {author} {\bibinfo {author} {\bibfnamefont {P.}~\bibnamefont
  {Xiong}}, \bibinfo {author} {\bibfnamefont {G.}~\bibnamefont {Xiao}},
  \bibinfo {author} {\bibfnamefont {J.~Q.}\ \bibnamefont {Wang}}, \bibinfo
  {author} {\bibfnamefont {J.~Q.}\ \bibnamefont {Xiao}}, \bibinfo {author}
  {\bibfnamefont {J.~S.}\ \bibnamefont {Jiang}}, \ and\ \bibinfo {author}
  {\bibfnamefont {C.~L.}\ \bibnamefont {Chien}},\ }\href {\doibase
  10.1103/PhysRevLett.69.3220} {\bibfield  {journal} {\bibinfo  {journal}
  {Physical Review Letters}\ }\textbf {\bibinfo {volume} {69}},\ \bibinfo
  {pages} {3220} (\bibinfo {year} {1992})}\BibitemShut {NoStop}%
\bibitem [{\citenamefont {Ye}\ \emph {et~al.}(1999)\citenamefont {Ye},
  \citenamefont {Kim}, \citenamefont {Millis}, \citenamefont {Shraiman},
  \citenamefont {Majumdar},\ and\ \citenamefont {Te\v{s}anovi\'{c}}}]{YKM99}%
  \BibitemOpen
  \bibfield  {author} {\bibinfo {author} {\bibfnamefont {J.}~\bibnamefont
  {Ye}}, \bibinfo {author} {\bibfnamefont {Y.~B.}\ \bibnamefont {Kim}},
  \bibinfo {author} {\bibfnamefont {A.~J.}\ \bibnamefont {Millis}}, \bibinfo
  {author} {\bibfnamefont {B.~I.}\ \bibnamefont {Shraiman}}, \bibinfo {author}
  {\bibfnamefont {P.}~\bibnamefont {Majumdar}}, \ and\ \bibinfo {author}
  {\bibfnamefont {Z.}~\bibnamefont {Te\v{s}anovi\'{c}}},\ }\href {\doibase
  10.1103/PhysRevLett.83.3737} {\bibfield  {journal} {\bibinfo  {journal}
  {Physical Review Letters}\ }\textbf {\bibinfo {volume} {83}},\ \bibinfo
  {pages} {3737} (\bibinfo {year} {1999})}\BibitemShut {NoStop}%
\end{thebibliography}%

\end{document}